\documentclass[journal=acsnano,manuscript=article]{achemso}

\usepackage[version=3]{mhchem} % Formula subscripts using \ce{}
\usepackage{booktabs}  % For better horizontal lines
\usepackage{hyperref}
\usepackage{xcolor}
\usepackage[final]{changes}

\author{Kayla~Fallon}
\email{kayla.fallon@glasgow.ac.uk}
\affiliation{SUPA School of Physics and Astronomy, University of Glasgow, Glasgow, G12 8QQ, United Kingdom}

\author{Reshma~Peremadathil-Pradeep}
\affiliation{Swiss Federal Laboratories for Material Science and 
Technology (Empa), Ueberlandstrasse 129, 8600 Duebendorf, Switzerland}

\author{Christopher~E.~A.~Barker}
\affiliation{National Physical Laboratory, Hampton Road,
Teddington, TW11 0LW, United Kingdom}
\alsoaffiliation{School of Physics and Astronomy, University of Leeds, Leeds, LS2 9JT, United Kingdom}

\author{Zoey~Tumbleson}
\affiliation{Department of Physics, University of California, Santa Cruz, CA 95064, USA}

\author{Emily~Darwin}
\affiliation{Swiss Federal Laboratories for Material Science and 
Technology (Empa), Ueberlandstrasse 129, 8600 Duebendorf, Switzerland}

\author{Andrea~Meo}
\affiliation{Department of Electrical and Information Engineering, Politecnico di Bari, 70125 Bari, Italy}

\author{Eloi~Haltz}
\affiliation{School of Physics and Astronomy, University of Leeds, Leeds, LS2 9JT, United Kingdom}
\alsoaffiliation{LSPM-CNRS, UPR3407, Sorbonne Paris Nord University, Villetaneuse, France}

\author{Benjamin~A.~Brereton}
\affiliation{School of Physics and Astronomy, University of Leeds, Leeds, LS2 9JT, United Kingdom}

\author{Trevor~Almeida}
\affiliation{SUPA School of Physics and Astronomy, University of Glasgow, Glasgow, G12 8QQ, United Kingdom}

\author{Colin~Kirkbride}
\affiliation{SUPA School of Physics and Astronomy, University of Glasgow, Glasgow, G12 8QQ, United Kingdom}

\author{Sara~Villa}
\affiliation{SUPA School of Physics and Astronomy, University of Glasgow, Glasgow, G12 8QQ, United Kingdom}
\alsoaffiliation{AREA Science Park, Padriciano 99, 34149 Trieste, Italy}

\author{Sophie~A.~Morley}
\affiliation{Advanced Light Source, Lawrence Berkeley National Laboratory, Berkeley, CA 94720, USA}

\author{Mario~Carpentieri}
\affiliation{Department of Electrical and Information Engineering, Politecnico di Bari, 70125 Bari, Italy}

\author{Riccardo~Tomasello}
\affiliation{Department of Electrical and Information Engineering, Politecnico di Bari, 70125 Bari, Italy}

\author{Hans~J.~Hug}
\affiliation{Swiss Federal Laboratories for Material Science and 
Technology (Empa), Ueberlandstrasse 129, 8600 Duebendorf, Switzerland}

\author{Christopher~H.~Marrows}
\affiliation{School of Physics and Astronomy, University of Leeds, Leeds, LS2 9JT, United Kingdom}

\author{Stephen~McVitie}
\affiliation{SUPA School of Physics and Astronomy, University of Glasgow, Glasgow, G12 8QQ, United Kingdom}%

\title[An \textsf{achemso} demo]
{Beyond conventional skyrmions in synthetic antiferromagnets}


\begin{document}

%\begin{tocentry}

%Some journals require a graphical entry for the Table of Contents. This should be laid out ``print ready'' so that the sizing of the text is correct.

%Inside the \texttt{tocentry} environment, the font used is Helvetica 8\,pt, as required by \emph{Journal of the American Chemical Society}.

%The surrounding frame is 9\,cm by 3.5\,cm, which is the maximum permitted for  \emph{Journal of the American Chemical Society} graphical table of content entries. The box will not resize if the content is too big: instead it will overflow the edge of the box.

%This box and the associated title will always be printed on a separate page at the end of the document.

% \end{tocentry}


\begin{abstract}
Magnetic skyrmions are topologically protected spin textures that can act as reconfigurable nanoscale information carriers. In synthetic antiferromagnets (SAFs), interlayer exchange coupling offers an additional control parameter beyond the interfacial Dzyaloshinskii-Moriya interaction (DMI) and magnetic anisotropy. Here, we engineer a SAF composed of two chemically distinct ferromagnets (CoB and CoFeB), in which the external magnetic field and interlayer exchange act asymmetrically on the sublattices. The competition of these effects, acting as a resultant effective-field, gives rise to two distinct skyrmion families in different field regimes. \replaced{In large}{At high} fields, conventional-polarity skyrmions nucleate, with core antiparallel to the external field, whereas \replaced{in smaller}{at lower} fields an inverse-polarity skyrmion state emerges as the effective-field reverses sign and almost saturates the CoFeB layers.  Return-point memory measurements confirm independent nucleation pathways for the two families. Using element-resolved x-ray magnetometry, correlative magnetic force and Lorentz transmission electron microscopies, and parameter-matched micromagnetic modelling, we show that all textures reside only in the CoFeB layers, which experience a \added{Ruderman–Kittel–Kasuya–Yosida (}RKKY\added{)}  exchange field originating from the CoB layers.  This effective-field method provides a robust route to programmable three-dimensional spin textures with controlled polarity in selected layers of a multilayer with potential for applications in skyrmion-based computing and spin-logic architectures.
\end{abstract}

\section{Introduction}

Magnetic skyrmions, nanoscale topologically non-trivial spin textures, have emerged as candidate information carriers for beyond-CMOS logic and memory, as well as for unconventional and neuromorphic computing~\cite{Nagaosa2013,Fert2017,Marrows2021,Zzvorka2019,Finocchio2024,Bourianoff2018,Song2020,Yokouchi2022,Raab2022,Sun2023,Beneke2024}. Following their first observation in bulk chiral magnets~\cite{Muehlbauer2009}, magnetic multilayers composed of heavy metal$_1$/ferromagnet/heavy metal$_2$ (HM1/FM/HM2) repetitions with interfacial perpendicular magnetic anisotropy (PMA) and Dzyaloshinskii–Moriya interaction (DMI) enabled skyrmion stabilisation and manipulation at room temperature in materials compatible with CMOS processing~\cite{Chen2015,Jiang2015,Woo2016,Boulle2016,Legrand2017}. 
Magnetic multilayers represent a versatile platform, offering multiple degrees of freedom to tune the skyrmion stability. When every FM layer is the same, skyrmions typically appear in these multilayers with their cores as regions of reverse magnetisation close to the field-saturated state~\cite{Soumyanarayanan2017}. They form vertical `tubes' of reversed magnetisation threading through every FM layer in the multilayer stack. Magnetostatic, and in some systems also \added{Ruderman–Kittel–Kasuya–Yosida (}RKKY\added{)} exchange coupling, ensures that the skyrmion cores align in every FM layer, while the competition with the DMI determines the local chirality: for small DMI (and negligble RKKY interlayer coupling), the magnetostatic field promotes a thickness-dependent chirality, leading to a `hybrid' skyrmion, whereas for large DMI, pure Néel skyrmions with uniform chirality are stabilised~\cite{LeGrand2018,Duong2020,Raftrey2024}.

Magnetic multilayers also allow for the combination of different materials along the vertical direction. Piling up alternating ferrimagnet/antiferromagnet within the same stack leads to the stabilisation of ferrimagnetic tubular skyrmions without the need for HMs~\cite{Xu2021}. Sandwiching a ferrimagnet between two HM1/FM/HM2 multilayers enables the co-existence of two skyrmion phases: a tubular skyrmion propagating throughout the sample and an incomplete skyrmion confined to the FM layers~\cite{Mandru2020NatComm}.
Furthermore, engineering the PMA profile expands the family of achievable magnetic textures. Tailored PMA gradients allow for the stabilisation of more exotic 3D textures, such as skyrmion cocoons~\cite{Grelier2022} or hopfions~\cite{Kent2021}, where the spin textures are confined to selected internal layers while the external layers remain perpendicularly magnetised.

%{\color{red}\newline
Synthetic antiferromagnets (SAFs)~\cite{Duine2018} are a particular class of magnetic multilayers in which FM layers are antiferromagnetically coupled across spacer layers by the \deleted{Ruderman--Kittel--Kasuya--Yosida (}RKKY\deleted{)} interaction~\cite{Parkin1991}. They were initially used to study giant magnetoresistance~\cite{Baibich1988} and have been widely applied as the reference layer in magnetic tunnel junction devices~\cite{Bandiera2010}. They enable control of exchange bias and antiferromagnetic (AFM) coupling obtained via engineering of the number of repetitions and thickness of the layers within the SAF, leading to the control of domain nucleation and reversal when the FM layers have PMA~\cite{Hellwig2002,Hellwig2003,Hauet2008,PhysRevB.109.134437}.

More recently, SAFs have gained a renewed attention for the stabilisation~\cite{Legrand2020,darwin2024antiferromagnetic} and manipulation~\cite{Dohi2019,Juge2022,Pham2024,Dohi2025} of skyrmions, where high speeds~\cite{Buettner2018,Pham2024} and a suppressed skyrmion Hall effect \cite{Zhang2016,Tomasello2017,Dohi2019} are achieved. These skyrmions exist as `tubes' of antiphase AFM order in the SAF background~\cite{Dohi2025}, and similarly to a conventional magnetic multilayer, every FM layer contains a skyrmion spin texture alternating from layer to layer.
However, the potential of SAFs in stabilising different topological textures remains largely unexplored. Competing RKKY coupling, DMI and chemical asymmetry may give rise to emergent spin textures beyond the well-established SAF skyrmion ‘tubes’, revealing a richer and more versatile form of topological order in these multilayers.

Here, we show that SAF multilayers with chemically distinct but magnetically compensated sublattices can support two families of skyrmions, accessed in different field ranges, with opposite core polarities and that exist only within a single sublattice. We studied a compensated SAF with two sublattices consisting of CoB and CoFeB, shown in Fig.~\ref{Fig1_structure_SQUID}a, so that the PMA constant $K_\mathrm{u}$ and DMI strength $D$, as well as the AFM RKKY coupling strength $J_\mathrm{RKKY}$ can be tuned independently. This allows the external field and the RKKY field to act unequally on the two sublattices, enabling two distinct skyrmion families to exist in different field windows within a single field sweep, with core polarities opposing (conventional-polarity skyrmion) or aligned with the external field direction (inverse-polarity skyrmion). Combining magnetic force microscopy (MFM), Lorentz differential phase contrast (DPC) scanning transmission electron microscopy (STEM), soft X-ray techniques, and quantitative full micromagnetic simulations, we unequivocally attribute the observed textures to the CoFeB sublattice only, while the CoB sublattice remains in a homogenous magnetisation state and does not form any spin textures. Return point memory measurements show that the two families of skyrmions have distinct nucleation pathways. This suggests the possibility of being able to design multilayer stacks incorporating SAFs in which skyrmions can appear only in selected layers and with a chosen polarity at a given external field. Our results pave the way for multilayer devices that can be useful as unconventional computing units in which reconfigurable functionality is performed not only by the number of 3D textures and their location, but also by their specific form and depth-dependent profile.

\section{Results}

\subsection{Materials platform and global magnetic response}

\begin{figure*}[t]
    %\centering
    \includegraphics[width=85mm]{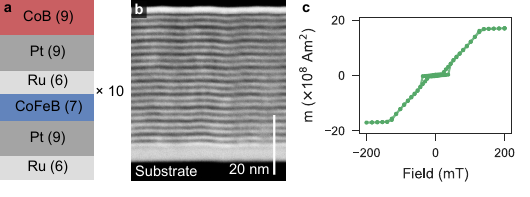}
    \caption{\textbf{
    SAF multilayer and compensated $m(H)$ response.} \textbf{a,} Schematic of a single repeat of the multilayer, Ta(50)/[Ru(6)/Pt(9)/CoFeB(7)/Ru(6)/Pt(9)/CoB(9)]$_{\times 10}$/Ru(6)/Pt(20); numbers are \added{rounded actual} layer thicknesses in \AA \added{ as determined from XRR, given in the methods section}. \textbf{b,} Cross-section ADF STEM image showing well-defined interfaces and high structural uniformity across repeats. \textbf{c,} Out-of-plane SQUID-VSM loop displaying a near-zero moment plateau around zero field (approximately $\pm 30$~mT), consistent with a fully compensated SAF.}
    \label{Fig1_structure_SQUID}
\end{figure*}

Fig.~\ref{Fig1_structure_SQUID} summarises the multilayer architecture and its compensated magnetic moment response. The schematic repeat unit is shown in Fig.~\ref{Fig1_structure_SQUID}a, with the CoFeB and CoB layer thicknesses chosen to give equal but opposite magnetic moments in the SAF. A representative cross-section annular dark-field (ADF) STEM image in Fig.~\ref{Fig1_structure_SQUID}b confirms interface quality and repeat uniformity. The out-of-plane superconducting quantum interference device-vibrating sample magnetometry (SQUID-VSM) loop in Fig.~\ref{Fig1_structure_SQUID}c exhibits a near-zero moment region (remanence is 1\% of $M_\mathrm{s}$) around zero field, consistent with a fully compensated SAF (more details are in the section Methods: Sample fabrication and metrology).

\subsection{Two skyrmion families revealed by correlative imaging}

\begin{figure*}[t!]
%\centering
\includegraphics[width=180mm]{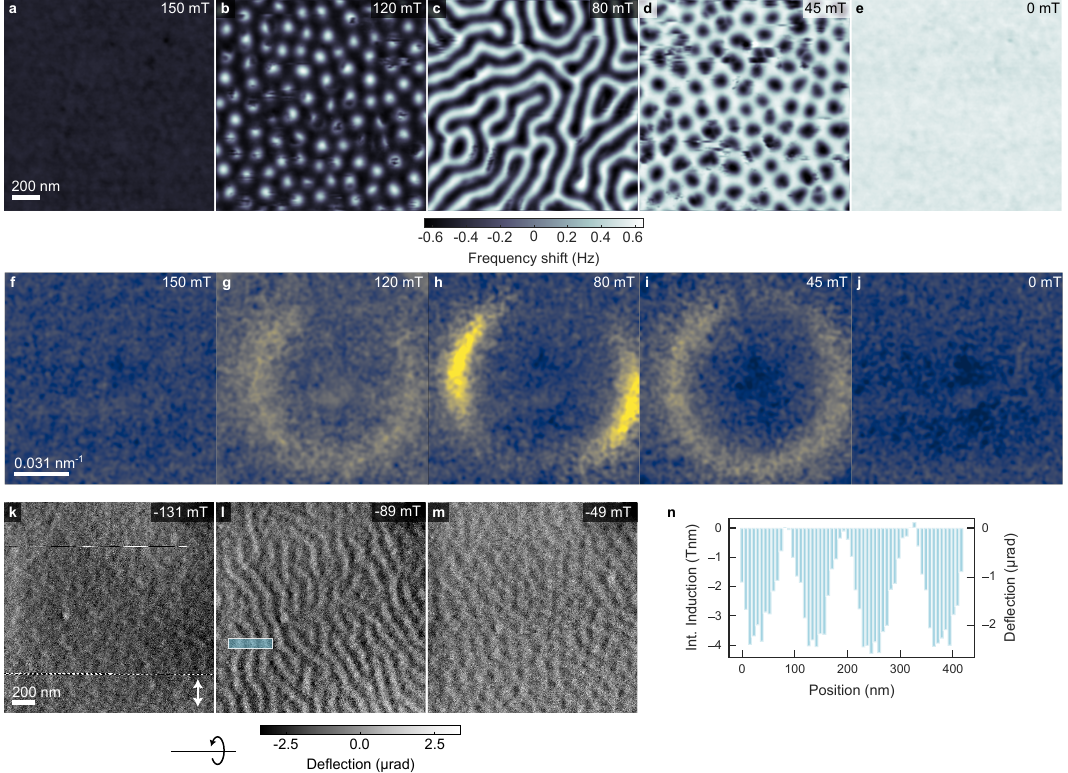}
\caption{\textbf{Correlative MFM, resonant SAXS and Lorentz STEM DPC imaging of the two skyrmion families.}
Top row (MFM, up-magnetised tip):
\textbf{a,} uniformly magnetised FM state at $+150$~mT;
\textbf{b,} conventional-polarity skyrmions with cores antiparallel to the external field (bright) at $+120$~mT;
\textbf{c,} maze-domains at $+80$~mT;
\textbf{d,} inverse-polarity skyrmions with cores parallel to the external field (dark) at $+45$~mT;
\textbf{e,} laterally uniform SAF state at $0$~mT.
Middle row, \textbf{f,}-\textbf{j,}: SAXS patterns at fields matched to MFM data.
Bottom row (Lorentz DPC STEM at matched states, fixed tilt $17^\circ$, and line profile):
\textbf{k,} conventional-polarity skyrmions at $-131$~mT;
\textbf{l,} maze-domains at $-89$~mT;
\textbf{m,} inverse-polarity skyrmions at $-49$~mT;
\textbf{n,} DPC line profile taken from the marked position in \textbf{g}. The DPC quantification is consistent with lateral variations confined to a single sublattice, while the counter-sublattice remains laterally uniform in the same field window.}
\label{Fig2_combined}
\end{figure*}

Correlative MFM, resonant small-angle x-ray scattering (SAXS) and Lorentz STEM experiments establish two distinct skyrmion families within the same field sweep (Fig.~\ref{Fig2_combined}). These techniques are complementary, being sensitive to $\mathbf{H}$, $\mathbf{M}$, and $\mathbf{B}$, respectively\added{, and even measuring different components of these quantities}. With an up-magnetised MFM tip, we observe a featureless saturated FM state at $+150$~mT (Fig.~\ref{Fig2_combined}a), a dense \replaced{large-field}{high-field} skyrmion state (conventional-polarity skyrmions) with cores antiparallel to the external field (bright contrast) at $+120$~mT (Fig.~\ref{Fig2_combined}b), a maze-domain state at $+80$~mT (Fig.~\ref{Fig2_combined}c), and a \replaced{small}{low}-field skyrmion state (inverse-polarity skyrmions) with inverted contrast (cores parallel to the external field (dark contrast) at $+45$~mT (Fig.~\ref{Fig2_combined}d). As the field is reduced to $0$~mT, the multilayer reaches a laterally uniform SAF state (Fig.~\ref{Fig2_combined}e). 

The corresponding reciprocal-space SAXS patterns (f–j) show \deleted{consistent} characteristic length scales for each texture, suggesting lateral dimensions below 100 nm, as detailed in Supplementary Note S3\added{, which is consistent with the visual size of features observed in the top and bottom row of Fig. \ref{Fig2_combined} (skyrmion diameters are discussed in more detail below)}.

Lorentz DPC STEM images, acquired at matched fields, corroborate the MFM results, showing induction signatures consistent with the same polarity reversal between the two skyrmion families (Fig.~\ref{Fig2_combined}k-m), with bright features in k and dark features in m. \added{Owing to the large fraction of non-magnetic material in the sample, and the predominantly out-of-plane magnetic texture, the DPC contrast is relatively weak. Accordingly, an extended field series, including images of saturated ferromagnetic and antiferromagnetic states is presented in Supplementary Note S2 to clarify the magnetic contribution.}  From the real-space images (MFM and DPC), we measure apparent skyrmion diameters well below 100~nm. The inverse-polarity skyrmions have diameters of \replaced{$81~\pm~4$~nm}{75–85~nm} in MFM and \replaced{$58~\pm~6$~nm}{50-65~nm} in DPC, while the conventional-polarity skyrmions measure \replaced{$64~\pm~4$~nm}{60–70~nm} in MFM at 120~mT and \replaced{$48~\pm~5$~nm}{44-55~nm} in DPC at -130~mT. The apparent diameter from the MFM measurements is expected to be slightly broader than the real skyrmion diameter, expansion on this (and further comments on these size measurements) are included in Supplementary Note S3.

DPC allows quantitative measurement of the integrated in-plane induction \cite{Krajnak2016} associated with the textures. In particular, Fig.~\ref{Fig2_combined}n shows the local integrated induction for the maze-domain state of Fig.~\ref{Fig2_combined}l. The average difference in the measured integrated induction is $3.9 \pm 0.2$~Tnm, which, after tilt correction (here, a specimen tilt is required to provide a non-zero $\mathbf{v}\times\mathbf{B}$ term in the Lorentz force \cite{Benitez2015,Fallon2019,Fallon2020,McVitie2018}), corresponds to an out-of-plane integrated induction of $12.8 \pm 0.7$~Tnm, in agreement with the value \replaced{($13.6 \pm 0.2$~Tnm)}{} calculated from SQUID-VSM magnetisation and x-ray reflectometry (XRR) thickness. This measurement is indicative of lateral transitions between vertical regions of full SAF alignment (grey) and full FM alignment (dark), and is therefore consistent with lateral variations confined to a single sublattice on top of a laterally uniform counter-sublattice. In other words, the DPC results suggest that the CoFeB and CoB sublattices exhibit two different magnetisation states. 

To unambiguously assign the textures to a specific sublattice and quantify each sublattice’s reversal, we therefore turn to element-specific transmission x-ray magnetic circular dichroism (XMCD) magnetometry in the next subsection.

\subsection{Element-resolved magnetometry and hysteresis loop reconstruction}

To disentangle the sublattice contributions and identify which layers host the field-evolving textures, we performed element-specific transmission XMCD magnetometry at normal incidence. Loops were recorded at the Co and Fe $L_3$ edges and probe, respectively, the combined CoB+CoFeB response and the CoFeB-only response, enabling reconstruction of the net global hysteresis loop and isolation of the CoB and CoFeB contributions.

\begin{figure*}[t]
    %\centering
    \includegraphics[width=180mm]{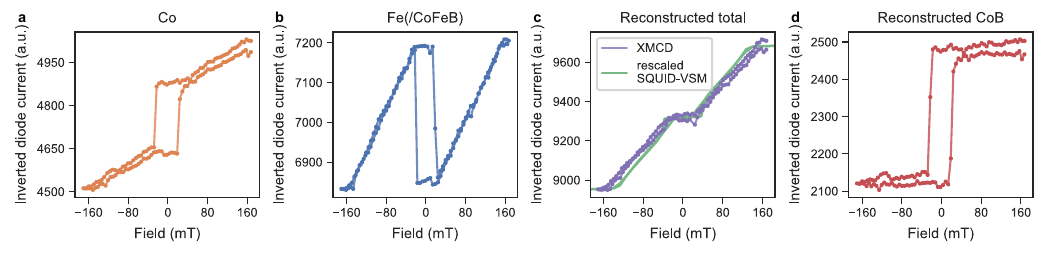}
    \caption{\textbf{Element-resolved XMCD out-of-plane loops and sublattice decomposition.} Transmission intensity $I(H)$ is linearly related to $m(H)$ for the selected element. 
    \textbf{a,} Co-tuned loop $I_\mathrm{Co}(H)$ (combined CoB+CoFeB response); 
    \textbf{b,} Fe-tuned loop $I_\mathrm{Fe}(H)$ (CoFeB-only); 
    \textbf{c,} reconstructed net loop $I_\mathrm{total}(H) = I_\mathrm{Fe}(H)+\left[I_\mathrm{Co}(H)-aI_\mathrm{Fe}(H)\right]$\added{ (purple)}, which reproduces the zero-moment plateau around $\mu_0 H=\pm 30$~mT observed by SQUID-VSM \replaced{(green)}{in Fig.~\ref{Fig1_structure_SQUID}\textbf{c}}; 
    \textbf{d,} isolated CoB signal $I_\mathrm{CoB}(H) = I_\mathrm{Co}(H)-aI_\mathrm{Fe}(H)$, which is a fully square loop. \added{The diode current has been inverted for comparison to SQUID-VSM data.}
    The scaling factor $a$ was determined as described in Methods.}
    \label{Fig2_SAXS_M-H-loops}
\end{figure*}

The Co-tuned loop (Fig.~\ref{Fig2_SAXS_M-H-loops}a) contains contributions from the CoB and CoFeB SAF layers but is dominated by the CoB sublattice because of its higher volume fraction of Co (9~nm Co$_{68}$B$_{32}$ vs.\ 7~nm Co$_{40}$Fe$_{40}$B$_{20}$). By contrast, the Fe-tuned loop (Fig.~\ref{Fig2_SAXS_M-H-loops}b) arises only from the CoFeB sublattice
and thus reveals exclusively its magnetic behaviour. Accordingly, we consider the element-resolved loop (transmitted intensity $I_\mathrm{Fe}$) as representing the sublattice-resolved loop of the CoFeB. Both loops exhibit a rectangular hysteretic central region, followed by a non-hysteretic linear increase or decrease of the magnetic moment for fields $\mu_0 H > 30$~mT or $\mu_0 H < -30$~mT, respectively.

To reconstruct the loop corresponding to the net out-of-plane magnetisation of the multilayer (i.e., the combined CoB + CoFeB contribution measured by the SQUID-VSM $m(H)$ loop in Fig.~\ref{Fig1_structure_SQUID}c), we introduce a scaling factor $a$ to account for the CoFeB contribution within the Co-tuned signal and define
\begin{equation}
    I_\mathrm{total}(H) = I_\mathrm{Fe}(H) + \left[ I_\mathrm{Co}(H) - a I_\mathrm{Fe}(H) \right].
\end{equation}
For a suitably chosen value of $a$ (see Methods), this composite loop (Fig.~\ref{Fig2_SAXS_M-H-loops}c)\added{, presented together with the SQUID-VSM data,} reproduces the zero-moment plateau observed by SQUID-VSM for $\mu_0 H$ between $\pm 30$~mT. We can now isolate the behaviour of the CoB sublattice by subtracting a scaled portion of the Fe signal (representing the CoFeB contribution) from the Co signal,
\begin{equation}\label{eq:SAXS_CoB}
    I_\mathrm{CoB}(H) = I_\mathrm{Co}(H) - a I_\mathrm{Fe}(H),
\end{equation}
which reveals a fully square loop (Fig.~\ref{Fig2_SAXS_M-H-loops}d) with rapid switching between negative and positive saturation at the edges of the SAF plateau at $\mu_0 H = \pm 30$~mT. The gradual change in $m(H)$ at \replaced{larger}{higher} fields, where spin textures are observed, only occurs in the loop for the CoFeB sublattice (Fig.~\ref{Fig2_SAXS_M-H-loops}b) under the influence of an effective field arising from the AFM interlayer coupling to the CoB sublattice. Therefore, the XMCD results confirm that the CoFeB sublattice hosts the field-evolving textures, whereas the CoB sublattice remains laterally uniform.

\subsection{Micromagnetic modelling}

To get further insights on the magnetic textures imaged in Fig.~\ref{Fig2_combined}, we performed full micromagnetic simulations of a SAF multilayer based on the experimental parameters (see Methods for details).
 
Relaxed configurations were obtained by field-cooling from randomised magnetisation in both CoB and CoFeB layers. We verified that alternative initial conditions (FM, AFM-aligned sublattices, and seeded skyrmions in every layer) relax to the same final states, indicating the robustness of the observed magnetic textures. 

\begin{figure*}[t]
    %\centering
    \includegraphics[width=180mm]{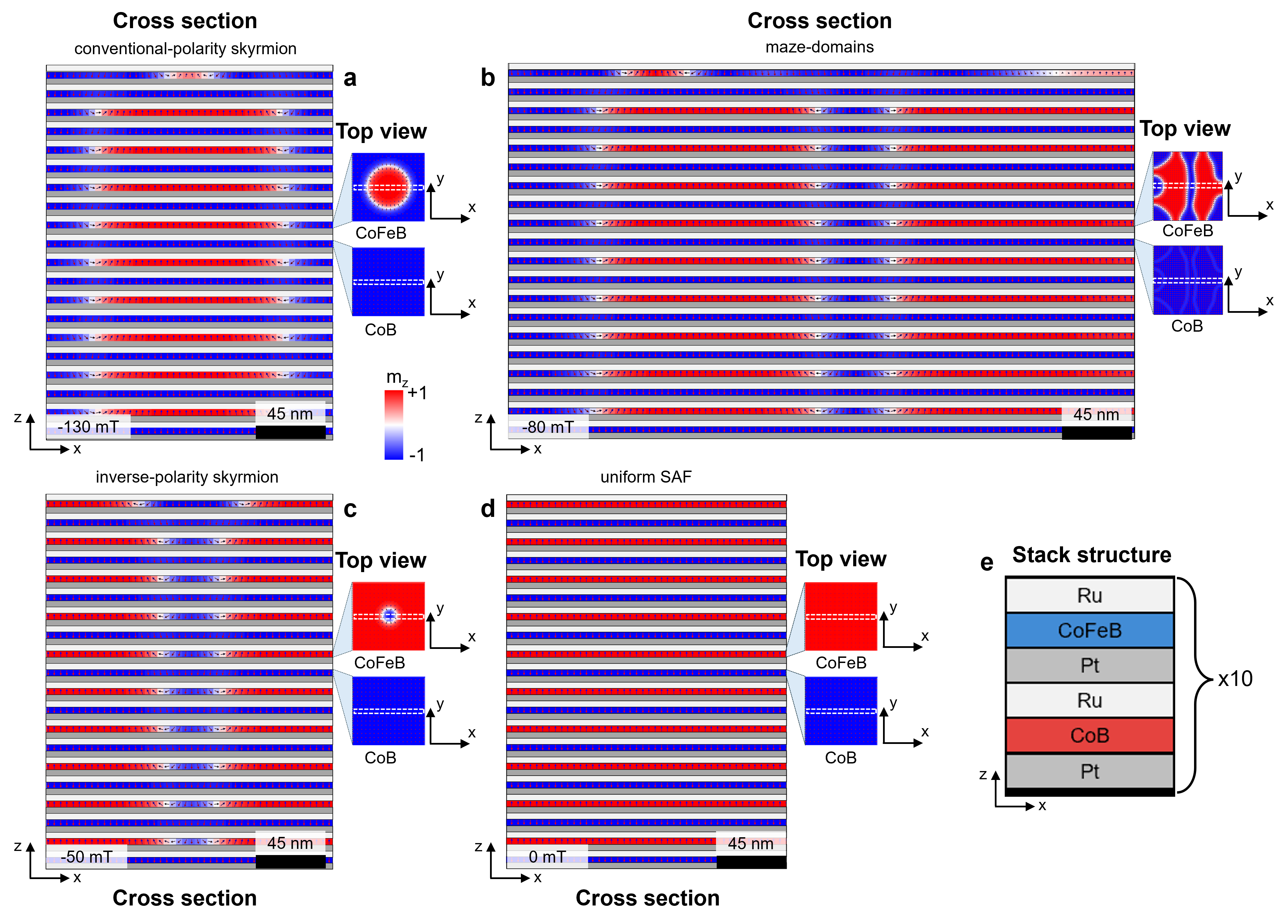}
    \caption{\textbf{Micromagnetic simulations results.}
    \textbf{a,} conventional-polarity skyrmions at $-130$~mT;
    \textbf{b,} maze-domains at $-80$~mT;
    \textbf{c,} inverse-polarity skyrmions at $-50$~mT;
    \textbf{d,} uniform SAF state at 0~mT.
    Textures form exclusively in the CoFeB layers. Scale and colour bars apply to all panels. 
    \textbf{e,} Sketch of the simulated multilayer stack. 
    }
    \label{fig:micro_relaxed_states}
\end{figure*}

Figure~\ref{fig:micro_relaxed_states} shows cross-section snapshots of the relaxed states at representative fields. Starting from \replaced{a large}{high} field, we obtain a conventional-polarity skyrmion state at $-130$~mT (Fig.~\ref{fig:micro_relaxed_states}a), a maze-domain state at $-80$~mT (Fig.~\ref{fig:micro_relaxed_states}b), an inverse-polarity skyrmion state at $-50$~mT (Fig.~\ref{fig:micro_relaxed_states}c), and a laterally uniform SAF state at 0~mT (Fig.~\ref{fig:micro_relaxed_states}d). Consistently with experiments, the skyrmion and maze-domain textures form exclusively in the CoFeB sublattice, whereas the CoB layers remain laterally uniform except for small perturbations in the vicinity of CoFeB domain walls. 
These localised perturbations mediate the coupling of adjacent CoFeB walls through CoB and help maintain the observed vertical coherence of the textures across repeats—consistent with the transmission measurements (SAXS and DPC).

An exception in the simulations appears in the very top CoFeB layer, where the skyrmion size deviates from that in the interior layers due to the asymmetric RKKY environment (CoB below, vacuum above). To isolate this effect, we repeated the simulations for a symmetric stack (CoB both at the bottom and the top), such that each CoFeB layer experiences similar interlayer coupling above and below. The top-layer asymmetry in skyrmion size disappears (not shown), confirming that interlayer coupling sets the depth profile of the texture and can bias the topmost layer. We point out that detecting such a single-layer deviation experimentally is challenging due to thickness integration, defects, pinning, and thermal effects.

Consistent with a sufficiently large interfacial DMI, the simulations show chiral N\'eel type textures across all states. The N\'eel nature was verified experimentally using Lorentz microscopy as shown in Supplementary Note S4.\added{ Therefore, we expect these skyrmions to exhibit a skyrmion Hall angle comparable to conventional ferromagnetic multilayers, since there is no cancellation of topological charge. Micromagnetic simulations of current-driven dynamics further confirm a finite skyrmion Hall angle (not shown). Because the chirality remains fixed while the core polarity reverses between the two skyrmion families, the associated topological charge changes sign, and the two skyrmion types therefore deflect in opposite transverse directions under current.} The simulations also show opposite trends of skyrmion size versus field in the two families, as shown in Supplementary Note S3.

\subsection{Return-point memory: two distinct nucleation mechanisms}

\begin{figure*}[t]
  %\centering
  \includegraphics[width=180mm]{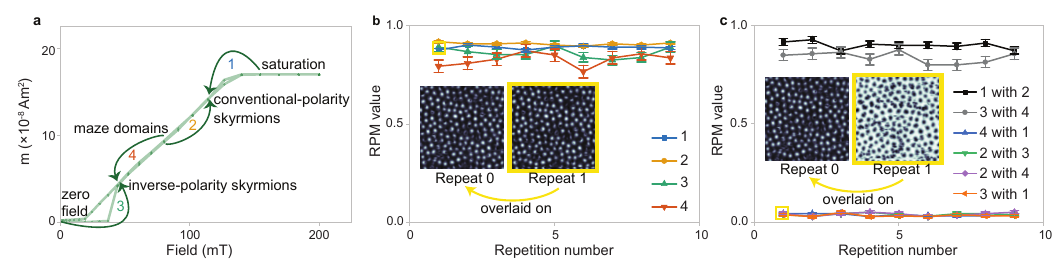}
    \caption{\textbf{Return-point memory experiments for the two skyrmion families.}
    \textbf{a,} Four nucleation paths are shown on the positive branch of the hysteresis loop: conventional-polarity skyrmions nucleated from magnetic saturation (1) or a maze-domain state (2); and inverse-polarity skyrmions nucleated from the zero-field SAF state (3) or maze-domain state (4).
    \textbf{b,} For each path, the initial skyrmion map (repeat 0) was overlaid with successive skyrmion maps (repeats 1 to 9) to calculate the RPM. Exemplary skyrmion maps from path 1 in the inset show that repeat 1 gets overlaid on repeat 0.
    \textbf{c,} Cross-type RPM analysis, in which an initial skyrmion map (repeat 0) nucleated via one path was overlaid with successive skyrmion maps (repeats 1 to 9) nucleated via different paths. Exemplary skyrmion maps from path 3 with path 1 in the inset show that path 3, repeat 1 gets overlaid on path 1, repeat 0.
    }
  \label{FIG8_3.pdf}
\end{figure*}

The MFM images revealed two distinct skyrmion families: conventional-polarity skyrmion (Fig. \ref{Fig2_combined}b) and inverse-polarity (Fig.\ref{Fig2_combined}d). To probe their thermodynamic stability and nucleation channels, we performed return point memory (RPM) experiments \cite{Baani2019,Katzgraber2006}. As shown in Fig.~\ref{FIG8_3.pdf}a, there are four possible skyrmion nucleation paths: conventional-polarity skyrmions nucleated either from saturation (path~1) or from a mid-field maze-domain state (path~2), and inverse-polarity skyrmions nucleated either from the zero-field SAF state (path~3) or from the maze-domain state (path~4). The external magnetic field was cycled ten times along each path. After each cycle, MFM was used to record skyrmion maps at the same external field ($+120$~mT for conventional-polarity skyrmions and $+40$~mT for inverse-polarity skyrmions). The successive skyrmion maps were overlaid with that from the initial cycle to determine the fraction of skyrmions reappearing at identical locations, defined as the RPM value. Details of the method for calculating the RPM values are given in Supplementary Note S5.

For both conventional-polarity and inverse-polarity skyrmions, the RPM analysis yielded consistently high values (0.8–0.9), regardless of the nucleation path (Fig.~\ref{FIG8_3.pdf}b), demonstrating robust re-nucleation at fixed pinning sites for a given skyrmion family. In contrast, cross-type RPM analysis (Fig.~\ref{FIG8_3.pdf}c) compares skyrmions nucleated via one path with skyrmions nucleated via another path. This analysis showed negligible RPM values (0.03–\replaced{0.05}{0.04}) when comparing the overlap between the conventional- and inverse-polarity skyrmions. However, it remained high (0.8-0.9) when comparing the same skyrmion family, nucleated via different paths.\added{ All RPM values are reported with standard deviation error bars, the errors range from $\pm$0.005 to $\pm$0.03.} This indicates uncorrelated nucleation sites between the conventional- and inverse-polarity skyrmions and hence, distinct pinning/formation mechanisms: external-field–driven for conventional-polarity skyrmions and RKKY–driven for inverse-polarity skyrmions.

\section{Discussion}%{Effective-field picture: external vs. RKKY competition}

\begin{figure*}[t]
  %\centering
  \includegraphics[width=180mm]{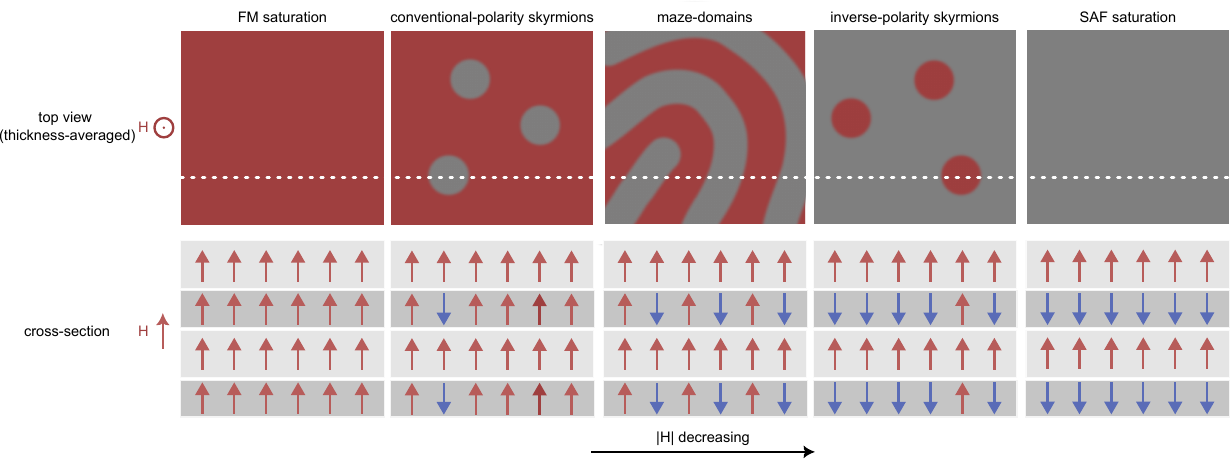}
  \caption{\textbf{Illustration of the effective-field mechanism generating two skyrmion families.}
  Top-down (thickness-averaged) and cross-section views of the SAF through the sequence
  \textbf{FM saturation} $\rightarrow$ \textbf{conventional-polarity skyrmions} $\rightarrow$ \textbf{maze-domains} $\rightarrow$ 
  \textbf{inverse-polarity skyrmions} $\rightarrow$ \textbf{uniform SAF state}.
  Grey indicates local antiferromagnetic alignment (zero net moment); red indicates locally ferromagnetic alignment.
  Skyrmion tubes form in CoFeB within a ferromagnetic background and reverse core polarity as the net effective field
  on CoFeB changes sign due to competition between the external field (up) and the antiparallel RKKY field imposed by
  the CoB sublattice across the 0.7~nm Ru spacers.}
  \label{Fig4_schematic}
\end{figure*}

The field-dependent evolution of the magnetisation states in the CoFeB sublattice can be qualitatively understood by considering the effective field acting on the CoFeB sublattice that arises from the RKKY coupling. At large positive external fields, the entire multilayer is uniformly magnetised in the upward direction (Fig.~\ref{Fig4_schematic}a). As the field is reduced, conventional-polarity N\'eel-type skyrmions appear in the CoFeB layers, with cores oriented antiparallel to the external field (Fig.~\ref{Fig4_schematic}b). These skyrmions extend coherently through the multilayer stack, forming AFM-aligned tubes embedded in an FM background. The effective field experienced by the CoFeB sublattice is governed by the competition between the positive external field and the opposing RKKY exchange field arising from the up-saturated CoB layers across the 0.7~nm Ru spacer.

With further reduction of the external field, the positive external field and the opposing RKKY field increasingly compensate each other, leading to the emergence of a maze-domain state (Fig.~\ref{Fig4_schematic}c). This marks a distinct behaviour from conventional skyrmion-hosting multilayers, where such a maze state would typically occur only at zero external field \cite{Soumyanarayanan2017, Raju2019, LeGrand2018}. When the external positive field is reduced further, the AFM RKKY field becomes dominant, corresponding to a net negative effective field acting on the CoFeB layers. As a result, domains with up magnetisation, which are antiparallel to the net effective field, shrink. As the external field is reduced further, the resulting RKKY negative field becomes sufficiently strong to nearly saturate the CoFeB layers, such that inverse-polarity skyrmions with up-core magnetisation develop (Fig.~\ref{Fig4_schematic}d). 

These skyrmions again propagate through the multilayer stack, but as FM-aligned tubes incorporated in an AFM background.  Finally, at zero external field, the strong AFM RKKY interaction saturates the CoFeB layers in the up state, driving the entire multilayer into the SAF ground state.

\added{The mechanism reported here reflects a general materials-design principle rather than a peculiarity of the present stack. It requires (i) strong antiferromagnetic interlayer exchange, (ii) global magnetic moment compensation, and (iii) engineered asymmetry in magnetic energy scales between the two sublattices.}

\added{In our system, moment compensation is achieved via a deliberate combination of material and thickness asymmetry between CoB and CoFeB. This not only balances the net magnetisation, but also generates distinct magnetic parameters in the two sublattices. In particular, CoFeB exhibits a substantially reduced perpendicular anisotropy $K_\mathrm{eff}$ (Table~1), placing it closer to the instability threshold where exchange, DMI and dipolar interactions favour non-collinear textures, whereas CoB remains robustly collinear. Although both interfaces contribute interfacial DMI, the relevant quantity is the DMI energy density $D_\mathrm{eff} = D/t$. The smaller thickness of CoFeB compensates for its potentially reduced intrinsic DMI (arising from the lower Co content at the CoFeB/Pt interface), yielding comparable $D_\mathrm{eff}$ in both layers; the decisive asymmetry therefore arises from the enhanced ratio $D_\mathrm{eff}/K_\mathrm{eff}$ in CoFeB.
}

\added{The strong RKKY coupling across the Ru spacer enforces antiparallel alignment and penalizes deviations from collinearity. Once a texture nucleates in the softer layer, the RKKY interaction acts as an effective exchange field on the harder one, suppressing texture formation there and confining the skyrmions to a single sublattice. Micromagnetic simulations confirm that reducing the RKKY strength restores texture formation in both layers, whereas stronger coupling localizes the non-collinear state.
This asymmetric SAF architecture therefore moves beyond conventional mirrored-texture concepts and enables layer-resolved phase coexistence and controllable three-dimensional spin-texture design. The strategy is general and transferable to compensated SAF platforms with tailored magnetic energy hierarchies.}

\section{Summary and Outlook}

\added{In summary, our multi-technique investigation of compensated synthetic antiferromagnets with chemically distinct sublattices reveals two skyrmion families—conventional- and inverse-polarity—with opposite core polarities that emerge in distinct magnetic-field windows and through different nucleation pathways. Remarkably, non-collinear textures localize exclusively within the magnetically softer CoFeB sublattice, while the CoB sublattice remains laterally uniform. This selective texture formation arises from a deliberately engineered asymmetry in magnetic energy scales, combined with strong antiferromagnetic interlayer exchange across the Ru spacers. The resulting states can be understood as a form of phase coexistence during the AFM–FM transition, in which tubular regions of one magnetic phase are embedded within the other.}

\added{Beyond the specific multilayer studied here, our results establish a general materials-design principle: by tailoring sublattice asymmetry in anisotropy, DMI and thickness under global moment compensation, texture formation can be confined to a single magnetic sublattice within an antiferromagnetically coupled stack. This expands the conventional SAF paradigm of mirrored textures and enables layer-resolved phase coexistence and programmable three-dimensional spin architectures. The ability to encode functionality not only in skyrmion number and position but also in their polarity and vertical profile introduces a new degree of freedom for complex magnetic heterostructures.}

\added{Looking forward, electrically driven control of skyrmion polarity represents a natural extension of this design strategy. Although uniform electric-field or current-driven switching across the present multilayer stack is limited by its repeated geometry, a reduced architecture with carefully balanced RKKY exchange, anisotropy and DMI could render the two polarities nearly degenerate. In such an energy-balanced regime, modest electrically induced anisotropy modulation, via voltage-controlled magnetic anisotropy or magneto-ionic effects, may enable deterministic, field-free polarity switching. More broadly, asymmetric SAFs with engineered energy hierarchies offer a versatile platform for reconfigurable three-dimensional spin-texture materials.}

\section{Methods}

\subsection{Sample fabrication and metrology}

The multilayer samples, composition Ta(50)/\allowbreak{}[Ru(\replaced{$6.03~\pm~0.05$}{})/\allowbreak{}Pt(\replaced{$8.99 \pm 0.06$}{})/\allowbreak{}CoFeB(\replaced{$6.69~\pm~0.07$}{})/\allowbreak{}Ru(\replaced{$6.03~\pm~0.05$}{})/\allowbreak{}Pt(\replaced{$8.99~\pm~0.06$}{})/\allowbreak{}CoB(\replaced{$8.69~\pm~0.07$}{})]$_{\times10}$ /\allowbreak{}Ru(\replaced{$6.03~\pm~0.05$}{})/\allowbreak{}Pt(20); numbers are layer thicknesses in \AA \replaced{ as determined from XRR}{}, were deposited by magnetron sputtering with a base pressure of $\sim 1 \times 10^{-8}$~m\deleted{B}\added{b}ar and an argon growth pressure of $4.43 \times 10^{-3}$~m\deleted{B}\added{b}ar. \added{A Ru spacer thickness of 6~\AA{} was selected to maximize the antiferromagnetic interlayer coupling, while the Pt interface promotes strong PMA. The asymmetric interfaces either side of each magnetic layer ensure a significant interfacial DMI. The CoB and CoFeB thicknesses were tuned to ensure net magnetic moment compensation in the SAF, such that the total magnetic moment of the CoFeB layers balances that of the CoB layers.} 

Several sister samples were deposited together. For x-ray reflectometry (XRR), SQUID-VSM and MFM measurements, the multilayer was deposited on a thermally-oxidised Si substrate with a nominal oxide thickness of 100~nm\added{, as specified by the substrate manufacturer}. For the LTEM measurements, 25~nm thick Si$_{3}$N$_{4}$ membranes on Si frames with a window size of 200~\textmu m were used, and similarly, for the soft x-ray measurements, 200~nm Si$_3$N$_4$ membranes on high-resistivity Si frames with a window size of 500~\textmu m were used. Thicknesses were determined by fitting X-ray reflectivity (XRR) measurements of the samples with GenX software\cite{Glavic2022}, using XRR measurements of calibration samples to inform initial parameters of the fits. XRR data and fits are provided in the supplementary note 1. Growth rates for all layers in the sample were between 0.01 and 0.04~nm/s, a deposition rate that allowed the growth of smooth layers over many repetitions. 

\added{Magnetic moment versus field loops of the CoB/CoFeB multilayer sample were measured with a Quantum Design MPMS-3 SQUID-VSM. The saturation magnetisations of the CoB and CoFeB layers were then obtained from their identical magnetic moment of 8.85$\cdot 10^{-8}$\,Am$^{2}$, the area of the sample used for magnetometry and the CoB and CoFeB layer thicknesses as determined by XRR}. The experimentally determined magnetic parameters of the two sub-lattices are given in Table \ref{Tab1_parameters}. 
% \added{The values of $M_\mathrm{s}$ are based on the SQUID-VSM measurements of the moment of the full multilayer and the layer thicknesses determined by XRR. The values of $K_\mathrm{u}$ are based on the difference of the integrals under in-plane and out-of-plane hysteresis loops for individual Pt/CoB/Ru and Pt/CoFeB/Ru trilayers grown in the same way as the full multilayer. }
The \replaced{$J_\mathrm{RKKY} $($= -0.034 \pm 0.002$\,mJ\,m$^{-2}$)}{} value is obtained from inserting the multilayer out-of-plane saturation field $\mu_0 H_\mathrm{s} = 125$~mT into the expression $J_\mathrm{RKKY} = \mu_0 H_\mathrm{s} M_\mathrm{s} t_\mathrm{FM} /2$, where $t_\mathrm{FM}$ is the ferromagnetic layer thickness. Note that the product $M_\mathrm{s} t_\mathrm{FM}$ is the same for both the CoB and CoFeB layers.

\begin{table}[bth]
    %\centering
    \caption{Experimentally determined magnetic parameters of the two sublattices in the compensated SAF multilayer. The interlayer exchange coupling strength is $J_\mathrm{RKKY} = -0.034 \pm 0.002$\,mJ\,m$^{-2}$. The effective anisotropy $K_\mathrm{eff}$ was determined from in-plane VSM measurements on separate CoB and CoFeB reference multilayers with identical thicknesses. The uniaxial anisotropy $K_\mathrm{u}$ was obtained from $K_\mathrm{eff}$ and the corresponding saturation magnetisation $M_\mathrm{s}$ values of the full CoB/CoFeB multilayer stack.}
    \label{Tab1_parameters}
    \begin{tabular}{cccccc}
    \toprule  % Booktabs for a nicer top line
    Sublattice & moment (Am$^2$) & Thickness (\AA) & $M_\mathrm{s}$ (MAm$^{-1}$) 
    & $K_\mathrm{eff}$ (MJm$^{-3}$) 
    & $K_\mathrm{u}$ (MJm$^{-3}$ )   \\
    \midrule
    CoB 
    & 8.85$\cdot 10^{-8}$ 
    & $8.69 \pm 0.07$ 
    & $0.63 \pm 0.01$ 
    & $0.22 \pm 0.02$ 
    & $0.47 \pm 0.02$   \\
    CoFeB 
    & 8.85$\cdot 10^{-8}$ 
    & $6.69 \pm 0.07$ 
    & $0.81 \pm 0.01$ 
    & $0.11 \pm 0.03$ 
    & $0.53 \pm 0.03$   \\
    \bottomrule
    \end{tabular}
\end{table}

\subsection{Soft X-ray techniques}

Soft X-ray synchrotron measurements were performed on the scattering branch of the Coherent Scattering and Microscopy (COSMIC) beamline 7.0.1.1 at the Advanced Light Source, Lawrence Berkeley National Laboratory. The X-ray photon energy was tuned to the Fe $L_3$ edge, $E = 707$~eV, to probe the CoFeB layers and the Co $L_3$ edge, $E = 778$~eV, to study (primarily) the CoB layers, although of course part of the signal was also generated by the CoFeB layers. Multilayer samples grown \deleted{in }\added{on} Si$_3$N$_4$ membranes were studied in a transmission geometry at normal incidence, with magnetic fields applied along the sample out-of-plane direction with an in-vacuum cryo-cooled electromagnet designed by Berkeley Laboratory~\cite{Scharfstein2016}. 

Element-resolved hysteresis loops were measured by setting a Si photodiode detector at the straight-through position and measuring the transmitted photon intensity at normal incidence with the elliptically polarised undulator set for a fixed photon helicity as a function of external field: the XMCD effect gives rise to an absorption of the beam that is proportional to the out-of-plane component of the magnetisation for the element selected by the choice of photon energy.

The scaling factor, $a = 0.35$, used to reconstruct the total loop and CoB loop, was determined empirically by adjusting $a$ until the reconstructed total loop matched the SQUID-VSM loop, with a closed central area. In principle, $a$ could be determined from the relative elemental composition and the absorption cross-sections. However, due to the complexity of accurately determining the latter, we have adopted the empirical approach. 

Resonant small angle x-ray scattering (SAXS) measurements were collected on a two-dimensional detector in the transmission gemeotry. The detector used was a Andor ikonL charge-coupled device (CCD) with a $2048 \times 2048$ pixel sensor area with pixel size $13.5 \times 13.5$~\textmu m$^{2}$ and a sample-to-detector distance, $d =0.93$~m. Each image was taken with exposure time of 0.6~s. The out-of-plane external magnetic field was varied for a fixed photon energy and helicity. Each presented SAXS pattern is the result of subtraction of the raw data from data of a saturated state to minimise the contrast associated with the beam stopper. These images were then low-pass filtered to remove high-frequency noise and enhance visibility of the scattering.

\subsection{Magnetic force microscopy}

The MFM measurements were performed using a home-built system operating under high vacuum conditions ($10^{-6}$ mbar) where an in situ perpendicular magnetic field of up to 380~mT can be applied. In vacuum, the mechanical quality factor ($Q$) of the cantilever is larger than $2 \times 10^5$. This improves the measurement sensitivity by a factor of 40 in comparison to that obtained under ambient conditions\cite{Feng2022}. For these experiments, an SS-ISC uncoated cantilever from Team Nanotec GmbH with a tip radius less than 5~nm was used. The tip was sputter-coated, first with a 2~nm Ta seed layer and then with a 6.5~nm Co layer to make them sensitive to magnetic fields. \added{The cantilever’s resonance frequency was experimentally determined to be 59.4 kHz.} A Zurich Instruments phase-locked loop (PLL) was used to drive the cantilever oscillation on resonance with an amplitude kept constant at 10~nm. Therefore, the measured signal is the frequency shifts which results from the derivative of the tip-sample interaction force. 
\added{The measurement was carried out in a single pass, non-contact mode with a constant tip-sample distance of roughly 20 nm was maintained using a capacitive frequency-modulated tip-sample distance control \cite{zhao2018magnetic}. The scan time per line was set to be 1~s.}

\subsection{Transmission electron microscopy}

All TEM imaging was conducted on a JEOL ARM 200cF equipped with a cold-field emission gun and CEOS probe aberration corrector, operated with an accelerating voltage of 200 kV. For the high-resolution imaging, a cross-section was extracted from the multilayer grown on a thermally-oxidised Si substrate using a FEI Helios dual beam Xe$^{+}$ dual-beam FIB SEM. The cross-section STEM images were collected using a JEOL ADF detector, with a pixel size of 0.12~nm.

For magnetic imaging (Fresnel and DPC), the microscope was operated in Lorentz mode, with the objective lens weakly excited to act as an external field source. The field values were calibrated by using a custom TEM holder incorporating a Hall sensor at the sample position. The textures were determined to be N\'eel type using tilt-series defocused Fresnel imaging \cite{Fallon2019}, presented in supplementary note 4, using a defocus of 1~mm. To image magnetic textures such as these, the sample must be tilted \cite{Fallon2019,Benitez2015,McVitie2018,Pandey2020, Jiang2019}, therefore the external field is angled with respect to the sample plane. All field values quoted for Lorentz images correspond to the out-of-plane component for the multilayer. 

DPC measures the deflection of the undiffracted beam, which can be converted into the integrated induction \cite{McVitie2015}. For out-of-plane domains, the signal is tilt-corrected by division by $\tan(\theta)$, where $\theta$ is the tilt angle~\cite{Fallon2019}, yielding an out of plane integrated induction reference value corresponding to $M_{s}$. Quantification was restricted to regions with domain walls perpendicular to the tilt axis, since the contrast depends on the angle between the electron beam and domain walls \cite{Benitez2015,McVitie2018}. Furthermore, in DPC, the deflection differences are quantitative; however, the absolute zero-deflection reference is unknown for continuous thin film samples. Here, the zero is set by using assumptions informed primarily by the XMCD $I(H)$ loops and ensuring consistency across the different images. For this reason, though, quantification is based on difference measurements taken from line profiles, rather than absolute values.

The DPC datasets were collected using a 50~\textmu m condenser aperture, giving a probe semi-angle of 0.88~mrad (probe size of 3.5~nm), and a camera length of \deleted{2000~cm }\added{20~m}. The datasets were collected on a MerlinEM 1R direct-electron detector, with a sampling pixel size of 7.4~nm. Deflection maps were obtained by an edge-detection and cross-correlation algorithm, \added{which }reduc\deleted{ing}\added{es} sensitivity to structural variations \added{in the sample} \cite{Krajnak2016}\added{.}\deleted{;} \added{The} \deleted{analysis }code \added{used for this analysis} is freely available \cite{pixDPCcode}.

\subsection{Micromagnetic modelling}

Micromagnetic simulations are performed by means of PETASPIN~\cite{Giordano2012}, a state-of-the-art CUDA-native in-house micromagnetic solver. The magnetic dynamics of the system is obtained integrating numerically the Landau-Lifshitz-Gilbert (LLG) equation by using the Adams-Bashforth scheme~\cite{Raimondo2022}: 
\begin{equation}\label{eq:LLG_micro}
\frac{\partial\mathbf{m}}{\partial \tau} =
- \frac{1}{1+\alpha_\mathrm{G}^2} \, \left[ \mathbf{m} \times \mathbf{h}_{\mathrm{eff}} +\alpha_\mathrm{G} \mathbf{m}\times(\mathbf{m} \times \mathbf{h}_{\mathrm{eff}}) \right] ,
\end{equation}
where $\mathbf{m}=\mathbf{M}/M_\mathrm{s}$ is the normalised magnetisation vector with $M_\mathrm{s}$ the saturation magnetisation, $\tau = \gamma_0 M_\mathrm{s} t$ is the dimensionless time, with $\gamma_0$ the gyromagnetic ratio, and $\alpha_\mathrm{G}$ is the Gilbert damping constant. $\mathbf{h}_{\mathrm{eff}} = \mathbf{H}_{\mathrm{eff}}/(\mu_0 M_\mathrm{s})$ is the normalised effective field in units of $\mu_0 M_\mathrm{s}$, where $\mu_0$ is the vacuum permeability, and includes exchange, anisotropy, interfacial DMI, magnetostatic, and external fields~\cite{Mandru2020NatComm}. 

The exchange field includes the RKKY contribution ($\mathbf{H}_{\mathrm{RKKY}}$), coupling the ferromagnetic layers via indirect exchange across the metallic spacers. $\mathbf{H}_{\mathrm{RKKY}}$ between two consecutive FM layers $i$ \& $j$ is given by~\cite{darwin2024antiferromagnetic}:
\begin{equation} \label{eq:RKKY_micro}
    \mathbf{H}_{\mathrm{RKKY},i(j)}= \frac{- J_\mathrm{RKKY} }{M_{s,i(j)} t_\mathrm{FM}} \mathbf{m}_{j(i)} ,
\end{equation}
where $t_\mathrm{FM}$ is the thickness of the discretisation cell, that here corresponds to the FM layer thickness, and $J_\mathrm{RKKY}$ is the RKKY coupling strength between the two FM layers. 
The interfacial DMI field is implemented as~\cite{Tomasello2014}:
\begin{equation}\label{eq:iDMI_micro}
    \mathbf{H}_{\mathrm{DMI}} = -\frac{2 D}{M_s} \left[ \mathbf{u}_z \left( \nabla \cdot \mathbf{m} \right) - \nabla m_z \right] \,,
\end{equation}
where $\mathbf{u}_z$ is the unit vector along the out-of-plane direction, $D$ is the DMI strength and $m_z$ is the normalised z-component of the magnetisation.
The other contributions to the effective fields are expressed in their standard formalisms.

\replaced{The experimental stack is simulated via 10 repetitions of a CoB layer and a CoFeB layer separated by two non-magnetic spacers (see Figure~\ref{fig:micro_relaxed_states}(e)). We consider a 400\,nm\,$\times$\,400\,nm film discretised in cells of in-plane dimension 4~nm~$\times$~4~nm and thickness 6\,\AA\, (same as the experimental value for CoFeB), which allows us to use the experimentally derived parameters for CoFeB (see Table~\ref{Tab1_parameters}). For the CoB layer, since it is modelled with the same thickness of CoFeB, we use the same $M_\mathrm{s}$ as CoFeB to have a compensated SAF, while rescaling $K_\mathrm{u}$ to maintain the experimental CoB $K_\mathrm{eff}$. 
We consider the same $J_{\mathrm{RKKY}}$ as in the experimental estimation. 
We fix $A$ to $8$\,pJ/m for both layers~\cite{Mohammadi2019,Guang2023,Liu2025} and consider the interfacial DMI as a free parameter. Notably, the obtained values are consistent with literature~\cite{Tacchi2017,Belmeguenai2017,Alshammari2021}. 
The full list of parameters employed in the micromagnetic simulations is summarised in Table~\ref{table:mag_parameters_micro}.
Simulations are performed at zero temperature.
}{[Pt(9)/Co$_{40}$Fe$_{40}$B$_{20}$(6)/Ru(7)/Pt(9)/Co$_{68}$B$_{32}$(8)/Ru(7)]$_{\times 10}$ (see Figure~\ref{fig:micro_relaxed_states}(e)) layers are simulated by 10 repetitions of a CoB layer and a CoFeB layer separated by two non-magnetic spacers. 
To be able to describe the CoFeB layer properly, We ch\deleted{o}ose to discretise the system into cells of in-plane  dimensions 4~nm~$\times$~4~nm and 0.6~nm height, which results in a stack composed of 58 layers in total, and consider a film 400~nm~$\times$~400~nm wide.

To parametrise $M_\mathrm{s}$, $K_\mathrm{u}$ and $J_\mathrm{RKKY}$ of the ferromagnetic layers, we rely on the experimentally derived parameters for CoFeB (Table~\ref{Tab1_parameters}) wherever possible. We choose 
meaning that we need estimate only $A$ and $D$. We rescale the CoB parameters in order to maintain the full compensation of the stack, as in the experiments. 
To determine $D$ and $J_\mathrm{RKKY}$, instead, we vary these parameters and choose those that reproduce the formation of skyrmions in the CoFeB layer and with a size comparable to experiments: $J_\mathrm{RKKY} = -0.4$~mJm$^{-2}$ and $D(\mathrm{CoB}) = 0.6$~mJm$^{-2}$, $D(\mathrm{CoB}) = 1.0$~mJm$^{-2}$ for CoB and CoFeB, respectively.
We can understand the weaker $J_{RKKY}$ than extracted from experiments by considering that the micromagnetic system is ideal, i.e. there are no defects and pinning sites. These can affect the coercivity and result in a stronger effective coupling than that extracted from hysteresis loops. Simulations are performed at zero temperature. The full list of parameters employed in the micromagnetic simulations is summarised in Table~\ref{table:mag_parameters_micro}.
}

\begin{table}[!h]
    \centering

    \caption{Magnetic parameters used in micromagnetic simulations for of each sub-lattice in the SAF multilayer.}
    \label{table:mag_parameters_micro}

    \setlength{\tabcolsep}{4pt}
    \begin{tabular}{ccccccc}
    \toprule  % Booktabs for a nicer top line
    Sublattice & $M_\mathrm{s}$ (MAm$^{-1}$) & $K_\mathrm{eff}$ (MJm$^{-3}$) & $K_\mathrm{u}$ (MJm$^{-3}$) & $J_\mathrm{RKKY}$ (mJm$^{-2}$) & $A$ (pJm$^{-1}$)  & $D$ (mJm$^{-2}$)\\
    \midrule
    CoB   & 0.81 & 0.22 & 0.63 & $-0.03$ & 8.00 & 1.00 \\
    CoFeB & 0.81 & 0.11 & 0.52 & $-0.03$ & 8.00 & 1.10 \\
    \bottomrule
    \end{tabular}
    \vspace{2mm} % Adds a little spacing before the note
    \parbox{0.9\linewidth}{
    %\footnotesize
    %$^{*}$ From experiments. $^{\dagger}$ Estimated.
}
\end{table}

\section{Data Availability}

Data associated with this work is available from the University of Glasgow: Enlighten Data repository under a CC-BY license at [doi will be inserted later].

%References

\bibstyle{naturemag}
\bibliography{bibliography_new}

@article{Scharfstein2016,
title = {An Endstation with Cryogenic Coils Contributing to a 0.5 Tesla Field and 30-400k Sample Thermal Control},
author = {Greg A. Scharfstein and Diego Arbelaez and and J.-Y. Jung},
journal = {Proc. 9th Mechanical Engineering Design of Synchrotron Radiation Equipment and Instrumentation Conference},
volume = {1}, 
pages = {396}, 
year = {2016}
}

@article{Duine2018,
    title = {Synthetic antiferromagnetic spintronics}, 
    author = {R. A. Duine and Kyung-Jin Lee and Stuart S. P. Parkin and M. D. Stiles},
    journal = {Nature Phys.}, 
    volume  = {14}, 
    pages = {217–219},
    year = {2018}
}

@article{Baibich1988,
title = {Giant Magnetoresistance of (001){Fe}/(001){Cr} Magnetic Superlattices},
author = {M. N. Baibich and J. M. Broto and A. Fert and F. Nguyen Van Dau and F. Petroff and P. Etienne and G. Creuzet and A. Friederich and J. Chazelas}, 
journal = {Phys. Rev. Lett.}, 
volume = {61}, 
pages = {2472},
year = {1988}
}

@article{Bandiera2010,
title = {Comparison of Synthetic Antiferromagnets and Hard Ferromagnets as Reference Layer in Magnetic Tunnel Junctions With Perpendicular Magnetic Anisotropy},
author = {S. Bandiera and R. C. Sousa and Y. Dahmane and C. Ducruet and C. Portemont and V. Baltz and S. Auffret and I. L. Prejbeanu and B. Dieny},
journal = {IEEE Magn. Lett.},
volume = {1},
pages = {3000204},
year = {2010}
}

@article{Buettner2018,
title = {Theory of isolated magnetic skyrmions: {F}rom fundamentals to room temperature applications},
author = {Felix B\"{u}ttner and Ivan Lemesh  and Geoffrey S. D. Beach}, 
journal = {Sci. Rep.}, 
volume = {8}, 
pages = {4464}, 
year = {2018} 
}

@article{Parkin1991,
title = {Systematic variation of the strength and oscillation period of indirect magnetic exchange coupling through the 3d, 4d, and 5d transition metals},
author = {S. S. P. Parkin},
journal = {Phys. Rev. Lett.},
volume = {67}, 
pages = {3598},
year = {1991}
}

@article{LeGrand2018,
title = {Hybrid chiral domain walls and skyrmions in magnetic multilayers},
author = {William Legrand and Jean-Yves Chauleau and Davide Maccariello and Nicolas Reyren and Sophie Collin and Karim Bouzehouane and Nicolas Jaouen and Vincent Cros and Albert Fert},
journal = {Science Adv.},
year = {2018},
volume = {4}, 
pages = {aat0415},
doi = {10.1126/sciadv.aat0415}
}

@article{Jiang2015,
title = {Blowing magnetic skyrmion bubbles},
author = {Wanjun Jiang and Pramey Upadhyaya and Wei Zhang and Guoqiang Yu and M. Benjamin Jungfleisch and Frank Y. Fradin and John E. Pearson and Yaroslav Tserkovnyak and Kang L. Wang and Olle Heinonen and Suzanne G. E. te Velthuis and Axel Hoffmann},
journal = {Science},
year = {2015},
volume = {349}, 
pages = {283-286},
doi = {10.1126/science.aaa1442}
}

@article{Yokouchi2022,
title = {Pattern recognition with neuromorphic computing using magnetic field–induced dynamics of skyrmions},
author = {Tomoyuki Yokouchi and Satoshi Sugimoto and Bivas Rana and Shinichiro Seki and Naoki Ogawa and Yuki Shiomi and Shinya Kasai and Yoshichika Otani}, 
journal = {Science Adv.}, 
year = {2022},
volume = {8}, 
pages = {abq5652},
doi = {10.1126/sciadv.abq5652}
}

@article{Beneke2024,
title = {Gesture recognition with Brownian reservoir computing using geometrically confined skyrmion dynamics},
author = {Grischa Beneke and Thomas Brian Winkler and Klaus Raab and Maarten A. Brems and Fabian Kammerbauer and Pascal Gerhards and Klaus Knobloch and Sachin Krishnia and Johan H. Mentink and Mathias Kl\"{a}ui},
journal = {Nature Commun.},
volume  = {15}, 
pages = {8103},
year = {2024}
}

@article{Marrows2021,
title = {Perspective on skyrmion spintronics},
author = {C. H. Marrows and K. Zeissler},
journal = {Appl. Phys. Lett.},
volume = {119}, 
pages = {250502},
year = {2021}
}

@article{Chen2015,
title = {Room temperature skyrmion ground state stabilized through interlayer exchange coupling},
author = {Gong Chen and Arantzazu Mascaraque and Alpha T. N'Diaye and Andreas K. Schmid},
journal = {Appl. Phys. Lett.},
volume = {106}, 
pages = {242404}, 
year = {2015},
doi = {10.1063/1.4922726}
}

@article{Nagaosa2013,
  title   = {Topological properties and dynamics of magnetic skyrmions},
  author  = {Nagaosa, Naoto and Tokura, Yoshinori},
  journal = {Nature Nanotechnology},
  volume  = {8},
  pages   = {899--911},
  year    = {2013},
  doi     = {10.1038/nnano.2013.243}
}

@article{Fert2017,
  title   = {Magnetic skyrmions: advances in physics and potential applications},
  author  = {Fert, Albert and Reyren, Nicolas and Cros, Vincent},
  journal = {Nature Reviews Materials},
  volume  = {2},
  pages   = {17031},
  year    = {2017},
  doi     = {10.1038/natrevmats.2017.31}
}

@article{Muehlbauer2009,
  title   = {Skyrmion lattice in a chiral magnet},
  author  = {M{\"u}hlbauer, Sebastian and Binz, Bernhard and Jonietz, Fabian and Pfleiderer, Christian and Rosch, Achim and Neubauer, Andreas and Georgii, Robert and B{\"o}ni, Peter},
  journal = {Science},
  volume  = {323},
  number  = {5916},
  pages   = {915--919},
  year    = {2009},
  doi     = {10.1126/science.1166767}
}

@article{Legrand2020,
  title   = {Room-temperature stabilization of antiferromagnetic skyrmions in synthetic antiferromagnets},
  author  = {Legrand, William and Maccariello, Davide and Ajejas, Fernando and Collin, S{\'e}bastien and Vecchiola, Alexis and Bouzehouane, Karim and Reyren, Nicolas and Cros, Vincent and Fert, Albert},
  journal = {Nature Materials},
  volume  = {19},
  pages   = {34--42},
  year    = {2020},
  doi     = {10.1038/s41563-019-0468-3}
}

@article{Dohi2019,
  title   = {Formation and current-induced motion of synthetic antiferromagnetic skyrmion bubbles},
  author  = {Dohi, Tomoyuki and DuttaGupta, Shouvik and Fukami, Satoshi and Ohno, Hideo},
  journal = {Nature Communications},
  volume  = {10},
  pages   = {5153},
  year    = {2019},
  doi     = {10.1038/s41467-019-13182-6}
}

@article{Pham2024,
  title   = {Fast current-induced skyrmion motion in synthetic antiferromagnets},
  author  = {Pham, V. T. and Sisodia, N. and Di Manici, I. and Urrestarazu-Larra{\~n}aga, J. and Bairagi, K. and Pelloux-Prayer, J. and Guedas, R. and Buda-Prejbeanu, L. D. and Auffret, S. and Locatelli, A. and Mente{\c{s}}, T. O. and Pizzini, S. and Kumar, P. and Finco, A. and Jacques, V. and Gaudin, G. and Boulle, O.},
  journal = {Science},
  volume  = {384},
  number  = {6693},
  pages   = {307--312},
  year    = {2024},
  doi     = {10.1126/science.add5751}
}

@article{Raab2022,
  title   = {Brownian reservoir computing realized using geometrically confined skyrmion dynamics},
  author  = {Raab, Kim and Brems, Mikkel A. and Beneke, Gero and Dohi, Tomoyuki and Roth{\"o}rl, J{\"o}rg and Kammerbauer, Florian and Mentink, J. H. and Kl{\"a}ui, Mathias},
  journal = {Nature Communications},
  volume  = {13},
  pages   = {6982},
  year    = {2022},
  doi     = {10.1038/s41467-022-34309-2}
}

@article{Sun2023,
  title   = {Experimental demonstration of a skyrmion-enhanced strain-mediated physical reservoir computing system},
  author  = {Sun, Y. and Lin, T. and Lei, N. and Chen, X. and Kang, W. and Zhao, Z. and Wei, D. and Chen, C. and Pang, S. and Hu, L. and Yang, L. and Dong, E. and Zhao, L. and Liu, L. and Yuan, Z. and Ullrich, A. and Back, C. H. and Zhang, J. and Pan, D. and Zhao, J. and Feng, M. and Fert, A. and Zhao, W.},
  journal = {Nature Communications},
  volume  = {14},
  pages   = {3434},
  year    = {2023},
  doi     = {10.1038/s41467-023-39207-9}
}

@article{Mandru2020NatComm,
  title   = {Coexistence of distinct skyrmion phases observed in hybrid ferromagnetic/ferrimagnetic multilayers},
  author  = {Mandru, A.-O. and Y{\i}ld{\i}r{\i}m, O. and Tomasello, R. and Heistracher, P. and Penedo, M. and Giordano, A. and Suess, D. and Hug, H. J.},
  journal = {Nature Communications},
  volume  = {11},
  pages   = {6365},
  year    = {2020},
  doi     = {10.1038/s41467-020-20025-2}
}

@article{Giordano2012,
title = {{Semi-implicit integration scheme for Landau–Lifshitz–Gilbert-Slonczewski equation}},
author = {Giordano, A. and Finocchio, G. and Torres, L. and Carpentieri, M. and Azzerboni, B.},
doi = {10.1063/1.3673428},
issn = {0021-8979},
journal = {Journal of Applied Physics},
month = {apr},
number = {7},
pages = {07D112},
volume = {111},
year = {2012}
}

@article{Raimondo2022,
author = {Raimondo, Eleonora and Saugar, Elias and Barker, Joseph and Rodrigues, Davi and Giordano, Anna and Carpentieri, Mario and Jiang, Wanjun and Chubykalo-Fesenko, Oksana and Tomasello, Riccardo and Finocchio, Giovanni},
doi = {10.1103/PhysRevApplied.18.024062},
journal = {Physical Review Applied},
month = {aug},
number = {2},
pages = {024062},
publisher = {American Physical Society},
title = {{Temperature-Gradient-Driven Magnetic Skyrmion Motion}},
url = {https://doi.org/10.1103/PhysRevApplied.18.024062 https://link.aps.org/doi/10.1103/PhysRevApplied.18.024062},
volume = {18},
year = {2022}
}

@article{Glavic2022,
    author = {A. Glavic and M. Bj\"orck}, 
    title = {{GenX 3: the latest generation of an established tool}},
    journal = {J. Appl. Cryst.},
    volume = {55},
    pages = {1063-1071},
    year = {2022},
    doi = {10.1107/S1600576722006653}
}

@article{Feng2022, 
year = {2022}, 
title = {{Magnetic force microscopy contrast formation and field sensitivity}}, 
author = {Feng, Y. and Vaghefi, P. Mirzadeh and Vranjkovic, S. and Penedo, M. and Kappenberger, P. and Schwenk, J. and Zhao, X. and Mandru, A.-O. and Hug, H.J.}, 
journal = {Journal of Magnetism and Magnetic Materials}, 
issn = {0304-8853}, 
doi = {10.1016/j.jmmm.2022.169073}, 
pages = {169073}, 
volume = {551}
}

@article{darwin2024antiferromagnetic,
  title={Antiferromagnetic interlayer exchange coupled {Co}$_{68}${B}$_{32}$/Ir/Pt multilayers},
  author={Darwin, Emily and Tomasello, Riccardo and Shepley, Philippa M and Satchell, Nathan and Carpentieri, Mario and Finocchio, Giovanni and Hickey, B J},
  journal={Sci. Rep.},
  volume={14},
  number={1},
  pages={95},
  year={2024},
  publisher={Nature Publishing Group UK London}
}

@article{Benitez2015,
   author = {M. J. Benitez and A. Hrabec and A. P. Mihai and T. A. Moore and G. Burnell and D. Mcgrouther and C. H. Marrows and S. McVitie},
   doi = {10.1038/ncomms9957},
   issn = {20411723},
   journal = {Nature Communications},
   month = {12},
   publisher = {Nature Publishing Group},
   title = {{Magnetic microscopy and topological stability of homochiral N\'eel domain walls in a Pt/Co/AlO x trilayer}},
   volume = {6},
   year = {2015}
}

@article{Fallon2019,
   author = {K. Fallon and S. McVitie and W. Legrand and F. Ajejas and D. Maccariello and S. Collin and V. Cros and N. Reyren},
   doi = {10.1103/PhysRevB.100.214431},
   issn = {24699969},
   issue = {21},
   journal = {Physical Review B},
   month = {12},
   publisher = {American Physical Society},
   title = {{Quantitative imaging of hybrid chiral spin textures in magnetic multilayer systems by Lorentz microscopy}},
   volume = {100},
   year = {2019}
}

@article{McVitie2018,
   author = {S. McVitie and S. Hughes and K. Fallon and S. McFadzean and D. McGrouther and M. Krajnak and W. Legrand and D. MacCariello and S. Collin and K. Garcia and N. Reyren and V. Cros and A. Fert and K. Zeissler and C. H. Marrows},
   doi = {10.1038/s41598-018-23799-0},
   issn = {20452322},
   issue = {1},
   journal = {Scientific Reports},
   month = {12},
   pmid = {29632330},
   publisher = {Nature Publishing Group},
   title = {{A transmission electron microscope study of N\'eel skyrmion magnetic textures in multilayer thin film systems with large interfacial chiral interaction}},
   volume = {8},
   year = {2018}
}

@article{McVitie2015,
   author = {S. McVitie and D. McGrouther and S. McFadzean and D. A. MacLaren and K. J. O'Shea and M. J. Benitez},
   doi = {10.1016/j.ultramic.2015.01.003},
   issn = {18792723},
   journal = {Ultramicroscopy},
   month = {5},
   pages = {57-62},
   publisher = {Elsevier B.V.},
   title = {Aberration corrected Lorentz scanning transmission electron microscopy},
   volume = {152},
   year = {2015}
}

@article{Fallon2020,
   author = {Kayla Fallon and Sean Hughes and Katharina Zeissler and William Legrand and Fernando Ajejas and Davide Maccariello and Samuel McFadzean and William Smith and Damien McGrouther and Sophie Collin and Nicolas Reyren and Vincent Cros and Christopher H. Marrows and Stephen McVitie},
   doi = {10.1002/smll.201907450},
   issn = {16136829},
   issue = {13},
   journal = {Small},
   month = {4},
   pmid = {32141234},
   publisher = {Wiley-VCH Verlag},
   title = {Controlled Individual Skyrmion Nucleation at Artificial Defects Formed by Ion Irradiation},
   volume = {16},
   year = {2020}
}

@article{Krajnak2016,
   author = {Matus Krajnak and Damien McGrouther and Dzmitry Maneuski and Val O'Shea and Stephen McVitie},
   doi = {10.1016/j.ultramic.2016.03.006},
   issn = {18792723},
   journal = {Ultramicroscopy},
   month = {6},
   pages = {42-50},
   publisher = {Elsevier B.V.},
   title = {Pixelated detectors and improved efficiency for magnetic imaging in STEM differential phase contrast},
   volume = {165},
   year = {2016}
}

@misc{pixDPCcode,
  author       = {Matus Krajnak},
  title        = {Pixelated DPC},
  howpublished = {\url{https://github.com/matkraj/PixelatedDPC}},
  note         = {GitHub repository, accessed 2 Oct 2025},
}

@article{Juge2022,
author = {Juge, Rom{\'{e}}o and Sisodia, Naveen and Larra{\~{n}}aga, Joseba Urrestarazu and Zhang, Qiang and Pham, Van Tuong and Rana, Kumari Gaurav and Sarpi, Brice and Mille, Nicolas and Stanescu, Stefan and Belkhou, Rachid and Mawass, Mohamad-Assaad and Novakovic-Marinkovic, Nina and Kronast, Florian and Weigand, Markus and Gr{\"{a}}fe, Joachim and Wintz, Sebastian and Finizio, Simone and Raabe, J{\"{o}}rg and Aballe, Lucia and Foerster, Michael and Belmeguenai, Mohamed and Buda-Prejbeanu, Liliana D. and Pelloux-Prayer, Johan and Shaw, Justin M. and Nembach, Hans T. and Ranno, Laurent and Gaudin, Gilles and Boulle, Olivier},
doi = {10.1038/s41467-022-32525-4},
journal = {Nature Communications},
month = {aug},
number = {1},
pages = {4807},
pmid = {35974009},
publisher = {Springer US},
title = {{Skyrmions in synthetic antiferromagnets and their nucleation via electrical current and ultra-fast laser illumination}},
url = {https://www.nature.com/articles/s41467-022-32525-4},
volume = {13},
year = {2022}
}

@article{Zhang2016,
author = {Zhang, Xichao and Zhou, Yan and Ezawa, Motohiko},
doi = {10.1038/ncomms10293},
issn = {2041-1723},
journal = {Nature Communications},
month = {jan},
number = {1},
pages = {10293},
publisher = {Nature Publishing Group},
title = {{Magnetic bilayer-skyrmions without skyrmion Hall effect}},
url = {http://dx.doi.org/10.1038/ncomms10293 https://www.nature.com/articles/ncomms10293},
volume = {7},
year = {2016}
}

@article{Tomasello2017,
author = {Tomasello, R. and Puliafito, V. and Martinez, E. and Manchon, A. and Ricci, M. and Carpentieri, M. and Finocchio, G.},
doi = {10.1088/1361-6463/aa7a98},
journal = {Journal of Physics D: Applied Physics},
publisher = {IOP Publishing},
title = {{Performance of synthetic antiferromagnetic racetrack memory: domain wall versus skyrmion}},
url = {https://iopscience.iop.org/article/10.1088/1361-6463/aa7a98},
volume = {50},
year = {2017}
}

@article{Hellwig2002,
author = {Hellwig, O. and Maat, S. and Kortright, J. B. and Fullerton, Eric E.},
doi = {10.1103/PhysRevB.65.144418},
journal = {Physical Review B},
month = {mar},
number = {14},
pages = {144418},
title = {{Magnetic reversal of perpendicularly-biased Co/Pt multilayers}},
url = {https://link.aps.org/doi/10.1103/PhysRevB.65.144418},
volume = {65},
year = {2002}
}

@article{Hellwig2003,
author = {Hellwig, Olav and Kirk, Taryl L. and Kortright, Jeffrey B. and Berger, Andreas and Fullerton, Eric E.},
doi = {10.1038/nmat806},
journal = {Nature Materials},
month = {feb},
number = {2},
pages = {112--116},
title = {{A new phase diagram for layered antiferromagnetic films}},
url = {https://www.nature.com/articles/nmat806},
volume = {2},
year = {2003}
}

@article{Hauet2008,
author = {Hauet, T. and G{\"{u}}nther, C. M. and Hovorka, O. and Berger, A. and Im, M.-Y. and Fischer, P. and Eim{\"{u}}ller, T. and Hellwig, O.},
doi = {10.1063/1.2961001},
journal = {Applied Physics Letters},
number = {4},
title = {{Field driven ferromagnetic phase nucleation and propagation in antiferromagnetically coupled multilayer films with perpendicular anisotropy}},
url = {https://pubs.aip.org/apl/article/93/4/042505/323971/Field-driven-ferromagnetic-phase-nucleation-and},
volume = {93},
year = {2008}
}

@article{Pandey2020,
   author = {Nisrit Pandey and Maxwell Li and Marc De Graef and Vincent Sokalski},
   doi = {10.1063/1.5130411},
   issn = {21583226},
   issue = {1},
   journal = {AIP Advances},
   month = {1},
   publisher = {American Institute of Physics Inc.},
   title = {Stabilization of coupled Dzyaloshinskii domain walls in fully compensated synthetic anti-ferromagnets},
   volume = {10},
   year = {2020}
}

@article{Dohi2025,
   author = {Takaaki Dohi and Mona Bhukta and Fabian Kammerbauer and Venkata Krishna Bharadwaj and Ricardo Zarzuela and Aakanksha Sud and Maria Andromachi Syskaki and Duc Minh Tran and Thibaud Denneulin and Sebastian Wintz and Markus Weigand and Simone Finizio and Jörg Raabe and Robert Frömter and Rafal E. Dunin-Borkowski and Jairo Sinova and Mathias Kläui},
   doi = {10.1038/s41467-025-63759-7},
   issn = {20411723},
   issue = {1},
   journal = {Nature Communications },
   month = {12},
   pmid = {41006294},
   publisher = {Nature Research},
   title = {Observation of a non-reciprocal skyrmion Hall effect of hybrid chiral skyrmion tubes in synthetic antiferromagnetic multilayers},
   volume = {16},
   year = {2025}
}

@article{Zzvorka2019,
   author = {Jakub Zázvorka and Florian Jakobs and Daniel Heinze and Niklas Keil and Sascha Kromin and Samridh Jaiswal and Kai Litzius and Gerhard Jakob and Peter Virnau and Daniele Pinna and Karin Everschor-Sitte and Levente Rózsa and Andreas Donges and Ulrich Nowak and Mathias Kläui},
   doi = {10.1038/s41565-019-0436-8},
   issn = {17483395},
   issue = {7},
   journal = {Nature Nanotechnology},
   month = {7},
   pages = {658-661},
   pmid = {31011220},
   publisher = {Nature Research},
   title = {Thermal skyrmion diffusion used in a reshuffler device},
   volume = {14},
   year = {2019}
}

@article{Finocchio2024,
doi = {10.1088/2399-1984/ad299a},
url = {https://doi.org/10.1088/2399-1984/ad299a},
year = {2024},
month = {mar},
publisher = {IOP Publishing},
volume = {8},
number = {1},
pages = {012001},
author = {Finocchio, Giovanni and Incorvia, Jean Anne C and Friedman, Joseph S and Yang, Qu and Giordano, Anna and Grollier, Julie and Yang, Hyunsoo and Ciubotaru, Florin and Chumak, Andrii V and Naeemi, Azad J and Cotofana, Sorin D and Tomasello, Riccardo and Panagopoulos, Christos and Carpentieri, Mario and Lin, Peng and Pan, Gang and Yang, J Joshua and Todri-Sanial, Aida and Boschetto, Gabriele and Makasheva, Kremena and Sangwan, Vinod K and Trivedi, Amit Ranjan and Hersam, Mark C and Camsari, Kerem Y and McMahon, Peter L and Datta, Supriyo and Koiller, Belita and Aguilar, Gabriel H and Temporão, Guilherme P and Rodrigues, Davi R and Sunada, Satoshi and Everschor-Sitte, Karin and Tatsumura, Kosuke and Goto, Hayato and Puliafito, Vito and Åkerman, Johan and Takesue, Hiroki and Ventra, Massimiliano Di and Pershin, Yuriy V and Mukhopadhyay, Saibal and Roy, Kaushik and Ting Wang, I- and Kang, Wang and Zhu, Yao and Kaushik, Brajesh Kumar and Hasler, Jennifer and Ganguly, Samiran and Ghosh, Avik W and Levy, William and Roychowdhury, Vwani and Bandyopadhyay, Supriyo},
title = {Roadmap for unconventional computing with nanotechnology},
journal = {Nano Futures}
}

@article{Song2020,
   author = {Kyung Mee Song and Jae Seung Jeong and Biao Pan and Xichao Zhang and Jing Xia and Sunkyung Cha and Tae Eon Park and Kwangsu Kim and Simone Finizio and Jörg Raabe and Joonyeon Chang and Yan Zhou and Weisheng Zhao and Wang Kang and Hyunsu Ju and Seonghoon Woo},
   doi = {10.1038/s41928-020-0385-0},
   issn = {25201131},
   issue = {3},
   journal = {Nature Electronics},
   month = {3},
   pages = {148-155},
   publisher = {Nature Research},
   title = {Skyrmion-based artificial synapses for neuromorphic computing},
   volume = {3},
   year = {2020}
}

@article{Bourianoff2018,
    author = {Bourianoff, George and Pinna, Daniele and Sitte, Matthias and Everschor-Sitte, Karin},
    title = {Potential implementation of reservoir computing models based on magnetic skyrmions},
    journal = {AIP Advances},
    volume = {8},
    number = {5},
    pages = {055602},
    year = {2018},
    month = {01},
    issn = {2158-3226},
    doi = {10.1063/1.5006918},
    url = {https://doi.org/10.1063/1.5006918},
    eprint = {https://pubs.aip.org/aip/adv/article-pdf/doi/10.1063/1.5006918/12865630/055602_1_online.pdf},
}

@article{Legrand2017,
   author = {William Legrand and Davide Maccariello and Nicolas Reyren and Karin Garcia and Christoforos Moutafis and Constance Moreau-Luchaire and Sophie Collin and Karim Bouzehouane and Vincent Cros and Albert Fert},
   doi = {10.1021/acs.nanolett.7b00649},
   issn = {15306992},
   issue = {4},
   journal = {Nano Letters},
   month = {4},
   pages = {2703-2712},
   pmid = {28358984},
   publisher = {American Chemical Society},
   title = {Room-Temperature Current-Induced Generation and Motion of sub-100 nm Skyrmions},
   volume = {17},
   year = {2017}
}

@article{Boulle2016,
   author = {Olivier Boulle and Jan Vogel and Hongxin Yang and Stefania Pizzini and Dayane De Souza Chaves and Andrea Locatelli and Tevfik Onur Menteş and Alessandro Sala and Liliana D. Buda-Prejbeanu and Olivier Klein and Mohamed Belmeguenai and Yves Roussigné and Andrey Stashkevich and Salim Mourad Chérif and Lucia Aballe and Michael Foerster and Mairbek Chshiev and Stéphane Auffret and Ioan Mihai Miron and Gilles Gaudin},
   doi = {10.1038/nnano.2015.315},
   issn = {17483395},
   issue = {5},
   journal = {Nature Nanotechnology},
   month = {5},
   pages = {449-454},
   pmid = {26809057},
   publisher = {Nature Publishing Group},
   title = {Room-temperature chiral magnetic skyrmions in ultrathin magnetic nanostructures},
   volume = {11},
   year = {2016}
}

@article{Woo2016,
   author = {Seonghoon Woo and Kai Litzius and Benjamin Krüger and Mi Young Im and Lucas Caretta and Kornel Richter and Maxwell Mann and Andrea Krone and Robert M. Reeve and Markus Weigand and Parnika Agrawal and Ivan Lemesh and Mohamad Assaad Mawass and Peter Fischer and Mathias Kläui and Geoffrey S.D. Beach},
   doi = {10.1038/nmat4593},
   issn = {14764660},
   issue = {5},
   journal = {Nature Materials},
   month = {5},
   pages = {501-506},
   pmid = {26928640},
   publisher = {Nature Publishing Group},
   title = {Observation of room-temperature magnetic skyrmions and their current-driven dynamics in ultrathin metallic ferromagnets},
   volume = {15},
   year = {2016}
}

@article{Jiang2019,
   author = {Wanjun Jiang and Sheng Zhang and Xiao Wang and Charudatta Phatak and Qiang Wang and Wei Zhang and Matthias Benjamin Jungfleisch and John E. Pearson and Yizhou Liu and Jiadong Zang and Xuemei Cheng and Amanda Petford-Long and Axel Hoffmann and Suzanne G.E. Te Velthuis},
   doi = {10.1103/PhysRevB.99.104402},
   issn = {24699969},
   issue = {10},
   journal = {Physical Review B},
   month = {3},
   publisher = {American Physical Society},
   title = {Quantifying chiral exchange interaction for Néel-type skyrmions via Lorentz transmission electron microscopy},
   volume = {99},
   year = {2019}
}

@article{Duong2020,
author = {Duong, Nghiep Khoan and Tomasello, Riccardo and Raju, M. and Petrovi{\'{c}}, Alexander P. and Chiappini, Stefano and Finocchio, Giovanni and Panagopoulos, Christos},
doi = {10.1063/5.0022033},
journal = {APL Materials},
month = {nov},
number = {11},
pages = {111112},
publisher = {AIP Publishing, LLC},
title = {{Magnetization reversal signatures of hybrid and pure N{\'{e}}el skyrmions in thin film multilayers}},
url = {https://doi.org/10.1063/5.0022033 http://aip.scitation.org/doi/10.1063/5.0022033 https://pubs.aip.org/apm/article/8/11/111112/569846/Magnetization-reversal-signatures-of-hybrid-and},
volume = {8},
year = {2020}
}

@article{Raftrey2024,
author = {David Raftrey  and Simone Finizio  and Rajesh V. Chopdekar  and Scott Dhuey  and Temuujin Bayaraa  and Paul Ashby  and Jörg Raabe  and Tiffany Santos  and Sinéad Griffin  and Peter Fischer },
title = {Quantifying the topology of magnetic skyrmions in three dimensions},
journal = {Science Advances},
volume = {10},
number = {40},
pages = {eadp8615},
year = {2024},
doi = {10.1126/sciadv.adp8615},
URL = {https://www.science.org/doi/abs/10.1126/sciadv.adp8615}
}

@article{Xu2021,
author = {Xu, Teng and Chen, Zhen and Zhou, Heng-An and Wang, Zidong and Dong, Yiqing and Aballe, Lucia and Foerster, Michael and Gargiani, Pierluigi and Valvidares, Manuel and Bracher, David M. and Savchenko, Tatiana and Kleibert, Armin and Tomasello, Riccardo and Finocchio, Giovanni and Je, Soong-Guen and Im, Mi-Young and Muller, David A. and Jiang, Wanjun},
doi = {10.1103/PhysRevMaterials.5.084406},
journal = {Physical Review Materials},
month = {aug},
number = {8},
pages = {084406},
publisher = {American Physical Society},
title = {{Imaging the spin chirality of ferrimagnetic N{\'{e}}el skyrmions stabilized on topological antiferromagnetic Mn3Sn}},
url = {https://link.aps.org/doi/10.1103/PhysRevMaterials.5.084406},
volume = {5},
year = {2021}
}

@article{Kent2021,
author = {Kent, Noah and Reynolds, Neal and Raftrey, David and Campbell, Ian T. G. and Virasawmy, Selven and Dhuey, Scott and Chopdekar, Rajesh V. and Hierro-Rodriguez, Aurelio and Sorrentino, Andrea and Pereiro, Eva and Ferrer, Salvador and Hellman, Frances and Sutcliffe, Paul and Fischer, Peter},
doi = {10.1038/s41467-021-21846-5},
issn = {2041-1723},
journal = {Nature Communications},
month = {mar},
number = {1},
pages = {1562},
pmid = {33692363},
publisher = {Springer US},
title = {{Creation and observation of Hopfions in magnetic multilayer systems}},
url = {https://www.nature.com/articles/s41467-021-21846-5},
volume = {12},
year = {2021}
}

@article{Grelier2022,
author = {Grelier, Matthieu and Godel, Florian and Vecchiola, Aymeric and Collin, Sophie and Bouzehouane, Karim and Fert, Albert and Cros, Vincent and Reyren, Nicolas},
doi = {10.1038/s41467-022-34370-x},
issn = {2041-1723},
journal = {Nature Communications},
month = {nov},
number = {1},
pages = {6843},
pmid = {36369167},
publisher = {Springer US},
title = {{Three-dimensional skyrmionic cocoons in magnetic multilayers}},
url = {https://www.nature.com/articles/s41467-022-34370-x},
volume = {13},
year = {2022}
}

@article{PhysRevB.109.134437,
  title = {Phase coexistence and transitions between antiferromagnetic and ferromagnetic states in a synthetic antiferromagnet},
  author = {Barker, C. E. A. and Fallon, K. and Barton, C. and Haltz, E. and Almeida, T. P. and Villa, S. and Kirkbride, C. and Maccherozzi, F. and Sarpi, B. and Dhesi, S. S. and McGrouther, D. and McVitie, S. and Moore, T. A. and Kazakova, O. and Marrows, C. H.},
  journal = {Phys. Rev. B},
  volume = {109},
  issue = {13},
  pages = {134437},
  numpages = {9},
  year = {2024},
  month = {Apr},
  publisher = {American Physical Society},
  doi = {10.1103/PhysRevB.109.134437},
  url = {https://link.aps.org/doi/10.1103/PhysRevB.109.134437}
}

@article{Baani2019,
   author = {Mirko Baćani and Miguel A. Marioni and Johannes Schwenk and Hans J. Hug},
   doi = {10.1038/s41598-019-39501-x},
   issn = {20452322},
   issue = {1},
   journal = {Scientific Reports},
   month = {12},
   pmid = {30816268},
   publisher = {Nature Publishing Group},
   title = {How to measure the local Dzyaloshinskii-Moriya Interaction in Skyrmion Thin-Film Multilayers},
   volume = {9},
   year = {2019}
}

@article{Katzgraber2006,
   author = {Helmut G. Katzgraber and Gergely T. Zimanyi},
   doi = {10.1103/PhysRevB.74.020405},
   issn = {10980121},
   issue = {2},
   journal = {Physical Review B},
   title = {Hysteretic memory effects in disordered magnets},
   volume = {74},
   year = {2006}
}

@article{Raju2019,
   author = {M. Raju and A. Yagil and Anjan Soumyanarayanan and Anthony K.C. Tan and A. Almoalem and Fusheng Ma and O. M. Auslaender and C. Panagopoulos},
   doi = {10.1038/s41467-018-08041-9},
   issn = {20411723},
   issue = {1},
   journal = {Nature Communications},
   month = {12},
   pmid = {30842413},
   publisher = {Nature Publishing Group},
   title = {The evolution of skyrmions in Ir/Fe/Co/Pt multilayers and their topological Hall signature},
   volume = {10},
   year = {2019}
}

@article{Soumyanarayanan2017,
   author = {Anjan Soumyanarayanan and M. Raju and A. L.Gonzalez Oyarce and Anthony K.C. Tan and Mi Young Im and A. P. Petrovic and Pin Ho and K. H. Khoo and M. Tran and C. K. Gan and F. Ernult and C. Panagopoulos},
   doi = {10.1038/NMAT4934},
   issn = {14764660},
   issue = {9},
   journal = {Nature Materials},
   month = {9},
   pages = {898-904},
   pmid = {28714983},
   publisher = {Nature Publishing Group},
   title = {Tunable room-temperature magnetic skyrmions in Ir/Fe/Co/Pt multilayers},
   volume = {16},
   year = {2017}
}

@article{Tomasello2014,
author = {Tomasello, R. and Martinez, E. and Zivieri, R. and Torres, L. and Carpentieri, M. and Finocchio, G.},
doi = {10.1038/srep06784},
issn = {2045-2322},
journal = {Scientific Reports},
month = {oct},
number = {1},
pages = {6784},
title = {{A strategy for the design of skyrmion racetrack memories}},
url = {https://www.nature.com/articles/srep06784},
volume = {4},
year = {2014}
}

@article{zhao2018magnetic,
  title={Magnetic force microscopy with frequency-modulated capacitive tip--sample distance control},
  author={Zhao, X and Schwenk, J and Mandru, AO and Penedo, M and Ba{\'c}ani, M and Marioni, MA and Hug, HJ},
  journal={New Journal of Physics},
  volume={20},
  number={1},
  pages={013018},
  year={2018},
  publisher={IOP Publishing}
}

@article{Tacchi2017,
author = {Tacchi, S. and Troncoso, R. E. and Ahlberg, M. and Gubbiotti, G. and Madami, M. and {\AA}kerman, J. and Landeros, P.},
doi = {10.1103/PhysRevLett.118.147201},
issn = {0031-9007},
journal = {Physical Review Letters},
month = {apr},
number = {14},
pages = {147201},
title = {{Interfacial Dzyaloshinskii-Moriya Interaction in Pt/CoFeB Films: Effect of the Heavy-Metal Thickness}},
url = {http://link.aps.org/doi/10.1103/PhysRevLett.118.147201},
volume = {118},
year = {2017}
}

@article{Belmeguenai2017,
author = {Belmeguenai, M. and Bouloussa, H. and Roussign{\'{e}}, Y. and Gabor, M. S. and Petrisor, T. and Tiusan, C. and Yang, H. and Stashkevich, A. and Ch{\'{e}}rif, S. M.},
doi = {10.1103/PhysRevB.96.144402},
issn = {2469-9950},
journal = {Physical Review B},
month = {oct},
number = {14},
pages = {144402},
title = {{Interface Dzyaloshinskii-Moriya interaction in the interlayer antiferromagnetic-exchange coupled Pt/CoFeB/Ru/CoFeB systems}},
url = {https://link.aps.org/doi/10.1103/PhysRevB.96.144402},
volume = {96},
year = {2017}
}

@article{Mohammadi2019,
author = {Mohammadi, Jamileh Beik and Kardasz, Bartek and Wolf, Georg and Chen, Yizhang and Pinarbasi, Mustafa and Kent, Andrew D.},
doi = {10.1021/acsaelm.9b00381},
issn = {26376113},
journal = {ACS Applied Electronic Materials},
number = {10},
pages = {2025--2029},
title = {{Reduced Exchange Interactions in Magnetic Tunnel Junction Free Layers with Insertion Layers}},
volume = {1},
year = {2019}
}

@article{Guang2023,
author = {Guang, Yao and Zhang, Like and Zhang, Junwei and Wang, Yadong and Zhao, Yuelei and Tomasello, Riccardo and Zhang, Senfu and He, Bin and Li, Jiahui and Liu, Yizhou and Feng, Jiafeng and Wei, Hongxiang and Carpentieri, Mario and Hou, Zhipeng and Liu, Junming and Peng, Yong and Zeng, Zhongming and Finocchio, Giovanni and Zhang, Xixiang and Coey, John Michael David and Han, Xiufeng and Yu, Guoqiang},
doi = {10.1002/aelm.202200570},
issn = {2199-160X},
journal = {Advanced Electronic Materials},
month = {jan},
number = {1},
pages = {2200570},
title = {{Electrical Detection of Magnetic Skyrmions in a Magnetic Tunnel Junction}},
url = {https://onlinelibrary.wiley.com/doi/10.1002/aelm.202200570},
volume = {9},
year = {2023}
}

@article{Liu2025,
author = {Liu, Shuhui and Tomasello, Riccardo and Wu, Yuxuan and Fang, Bin and Chen, Aitian and Zheng, Dongxing and Zhang, Baoshun and Darwin, Emily and Hug, Hans J and Carpentieri, Mario and Jiang, Wanjun and Zhang, Xixiang and Finocchio, Giovanni and Zeng, Zhongming},
doi = {10.1002/aelm.202500130},
issn = {2199-160X},
journal = {Advanced Electronic Materials},
month = {sep},
number = {15},
pages = {e00130},
title = {{Topological Skyrmion‐Based Spin‐Torque‐Diode Effect in Magnetic Tunnel Junctions}},
url = {https://advanced.onlinelibrary.wiley.com/doi/10.1002/aelm.202500130},
volume = {11},
year = {2025}
}

@article{Alshammari2021,
author = {Alshammari, Khulaif and Haltz, Eloi and Alyami, Mohammed and Ali, Mannan and Keatley, Paul S. and Marrows, Christopher H. and Barker, Joseph and Moore, Thomas A.},
doi = {10.1103/PhysRevB.104.224402},
journal = {Physical Review B},
month = {dec},
number = {22},
pages = {224402},
publisher = {American Physical Society},
title = {{Scaling of Dzyaloshinskii-Moriya interaction with magnetization in Pt/Co(Fe)B/Ir multilayers}},
url = {https://link.aps.org/doi/10.1103/PhysRevB.104.224402},
volume = {104},
year = {2021}
}

\begin{acknowledgement}

KF and SMcV were supported by EPSRC grant EP/T006811/1. The work of RT, MC and AM was supported by the projects PRIN20222N9A73 ``SKYrmion-based magnetic tunnel junction to design a temperature SENSor—SkySens'', Prin2022SAYARY “Metrology for spintronics: A machine learning approach for the reliable determination of the Dzyaloshinskii-Moriya interaction (MetroSpin)” funded by the Italian Ministry of Research and by the project PE0000021, ``Network 4 Energy Sustainable Transition - NEST'', funded by the European Union - Next Generation EU, under the National Recovery and Resilience Plan (NRRP), Mission 4 Component 2 Investment 1.3 - Call for Tender No. 1561 dated 11.10.2022 of the Italian MUR (CUP C93C22005230007).
CHM was supported by EPSRC grant EP/T006803/1.\added{ RPP, ED and HJH were supported through the SNF Lead Agency project 200021E 211828 ``Binary information represented by chiral spin textures''.} This research used resources of the Advanced Light Source, a U.S. DOE Office of Science User Facility under contract no. DE-AC02-05CH11231.

\end{acknowledgement}

\section{Author Information}

\subsection{Contributions}

K.F., C.K. and S.V. performed the Lorentz microscopy; K.F. analysed the data. 
R.P.P. and E.D. performed the magnetic force microscopy. 
C.E.A.B. and E.H. grew the multilayer stacks. C.E.A.B., E.H. and B.A.B. performed the SQUID-VSM measurements. B.A.B. carried out the XRR measurements. 
Z.T., S.A.M. and C.H.M. performed the soft x-ray measurements; Z.T. and K.F. analysed the data. 
T.A. prepared the cross-section TEM sample and performed the ADF STEM.
A.M., M.C. and R.T. performed the micromagnetic calculations and analysis. 
C.H.M. and S. McV. conceived the project, which was supervised by S.A.M., R.T., H.J.H., C.H.M., and S.McV.
K.F. wrote the initial draft of the manuscript. All authors reviewed and contributed to the manuscript prior to submission. 

\section{Ethics declarations}

\subsection{Competing interests}

The authors declare no competing interests.

\end{document}